\documentclass[]{aa}
\usepackage[utf8]{inputenc}
\usepackage{amsmath}
\usepackage{xcolor}
\usepackage{multirow}
\usepackage{xspace}
\usepackage{xstring}
\usepackage{supertabular}
\usepackage{tabularx} 
\usepackage{orcidlink} 
\usepackage{placeins}
\usepackage{hyperref}
\usepackage{soul}
\hypersetup{colorlinks=true, linkcolor=blue, filecolor=magenta, urlcolor=cyan, citecolor=blue}
\usepackage[varg]{txfonts}
\usepackage{caption}
\usepackage{subcaption}

\usepackage{placeins}

\begin{document} 

\title{Deformation of the CME and CME-driven shock due to interaction with the ambient solar wind: I. Modelling with Cone Model}

\author{A. Niemela
\inst{\ref{umbc}, 
\ref{nasa},\orcidlink{0000-0002-3746-9246}}\thanks{These authors contributed equally to this work.}
\and A. Valentino\inst{\ref{cmpa}\,\orcidlink{0000-0001-9702-468X}}\footnotemark[1]
\and J. Magdalenic\inst{\ref{cmpa},
{\ref{rob}}\,\orcidlink{0000-0003-1169-3722}}}
\institute{Goddard Planetary Heliophysics Institute, University of Maryland, Baltimore County, Baltimore, MD 21250, USA \label{umbc}
\and Heliospheric Physics Laboratory, Heliophysics Division, NASA Goddard Space Flight Center, Greenbelt, MD 20771, USA \label{nasa}
\and Center for Mathematical Plasma Astrophysics, KU Leuven, Leuven 3000, Belgium \label{cmpa}
\and Royal Observatory of Belgium, 1180 Ukkel, Brussels, Belgium \label{rob}}

\date{Received xx/xx/2025 Accepted xx/xx/2025}

\abstract
{The near-Earth environment is continuously impacted by the solar wind and the solar transients embedded in the wind. The largest eruptions of plasma and the magnetic field at the Sun are Coronal Mass Ejections (CMEs), which, if they arrive at Earth, can induce geomagnetic storms and negatively affect our technologically advanced societies.}
{This study aims to improve our understanding of the CME and the CME-driven shock wave interactions with the ambient solar wind. And further, how the evolution of the CME and the shock wave deformations in both time and space affect the space weather forecasting accuracy.}
{We studied two Earth-directed CMEs observed on December 7, 2020 and on October 28, 2021. For modelling of the solar wind and CME propagation in the inner heliosphere, we used the state-of-the-art 3D~MHD model EUHFORIA and the cone CME model. We focused our study on the regions along the main direction of propagation of CMEs and along Sun-Earth line.}
{Our results show that the deformations of the CME and the CME-driven shock can be significant, with the strongest impact out of the CME's main direction of propagation. Both studied CMEs propagate in a variable background solar wind, which results in different structuring of the CME and CME-driven shocks at very close angular distances. The first CME propagates through a mildly structured background wind, which results in the stronger deformation found predominantly further away from the main direction of propagation. On the other hand, the second studied CME propagates in a more complex environment, resulting in a strongly structured shock at very different locations, including the regions close to the main direction of propagation.}
{The interactions between the CME and the ambient solar wind in which they propagate also affect the characteristics of the CME-driven shock (e.g., the gas compression ratio across the shock). CMEs observed as the flank encounters at Earth can be strongly affected by the ambient solar wind conditions, with the expected differences in the arrival time at Earth reaching up to 16 hours.}

\titlerunning{Deformation of the CME-driven shocks}
\authorrunning{A. Niemela, A. Valentino \& J. Magdalenić.}
\maketitle

\section{Introduction}
\label{sect:introduction}

In recent decades, space weather science and operational space weather forecasting have become essential in protecting the critical infrastructure and enabling the reliable operation of space-borne and ground-based technologies. Coronal mass ejections (CMEs) are large expulsions of magnetic field and plasma from the Sun, and the most important space weather phenomena capable of inducing strong geomagnetic storms. High-speed solar wind streams, emanating from coronal holes, also play a crucial role in continuously modulating the heliospheric environment, as they can also trigger geomagnetic activity. Therefore, the improvements in the accuracy of forecasting the CME and solar wind properties as they propagate away from the Sun and upon their impact with Earth are a key goal of space weather research. 
    
Fast CMEs are capable of driving shock waves, which can be observed in white light, in situ data, or through the energetic particles that they accelerate and their associated radio emission. The shock-associated radio emission, so-called type-II radio bursts \citep{1950Wild,2006Mann,2010Magdalenic} are the oldest known signature of the shock waves. The frequency drift of the type-II bursts can be used to estimate the CME-driven shock velocity, which then can approximate the arrival of the CME and CME-driven shock at any place in the inner heliosphere \citep[see e.g., ][]{2010Vourlidas,2014Subramanian,Magdalenic14, Jebaraj20}. However, such estimation of the arrival time does not account for any interactions with preceding CMEs or for differences in the properties of the ambient solar wind, both of which could affect the CME and the shock propagation in the heliosphere \citep[see e.g., ][]{2024Wu}. Coronagraph observations, heliospheric imagers, and in situ measurements provide clear evidence that as CMEs propagate through the interplanetary medium, when we also name them ICMEs (interplanetary coronal mass ejections), they can undergo through the strong deformations. However, as the ICMEs expand into the inner heliosphere, they become more tenuous and harder to detect \citep{2012Harrison,2014Temmer,2020Jones,2021Chi}. Investigating ICME deformations through in situ data can also be very challenging, as these observations provide very limited spatial coverage and thus additional assumptions over the shape of the CME front during its transit in the heliosphere are required \citep{1999Sheeley,2006Owens,2012Davies}. In particular, \citet{2021Davies} focuses on an event on April 20, 2020, seen by multiple spacecraft in close proximity. The presented in situ data show very similar magnetic field signatures registered by BepiColombo, Solar Orbiter, and Wind. 
On the other hand, \citet{2024Palmerio} presents in situ data for the event on February 15, 2022, where BepiColombo and Parker Solar Probe (PSP) registered very different plasma characteristics despite being separated by only 0.03~au in the radial direction, 1.9$^{\circ}$ in latitude, and 4.3$^{\circ}$ in longitude. Such observations indicate that the background solar wind in which the CMEs are embedded plays a fundamental role in their propagation through the interplanetary medium \citep{2001Gopalswamy,2008Case,2011Temmer,2014Wang,2021Winslow}. During the solar minimum, the solar wind generally shows a more uniform structure, while the background wind during the maximum of the solar activity is often observed to be strongly structured. The strongly bimodal solar wind originating from the close to the CME source regions can induce the concave CME fronts as shown in \citet{2006Owens,2010Savani,2021Davies,2022Braga}. Additionally, the solar wind can be strongly disturbed by the preceding eruptions or blob-like structures in the solar wind sometimes observed in situ \citep[e.g., ][]{2023Niemela,2024Mayank}. Lower-density regions, i.e., solar wind streams, can significantly influence the propagation speed of different parts of the same CME \citep{2014Rollett}. Thus, to fully understand the CMEs' 3D structure and their interactions, 3D modelling is needed. 

Our current state-of-the-art forecasting tools are 3D magnetohydrodynamic (MHD) models, such as WSA-Enlil \citep{2004Odstrcil}, SUSANOO-CME \citep{2016Shiota}, SWASTi-CME \citep{2022Mayank} and EUHFORIA \citep{Pomoell2018, 2020Poedts}. However, it is often challenging to constrain the values of the CME input parameters, needed for the modelling \citep[see e.g., ][]{2024Valentino}. To obtain those values, white light coronagraph observations are most often employed by CME reconstruction tools. The majority of tools assume a simple croissant-shaped CME, similar to the cone CME model \citep{2004Xie}, which is, due to its simplicity, the most widely used CME model, particularly in operational forecasting. However, the observations-based CME reconstruction models often require approximations and simplifications, incorporating in such a way uncertainties into the MHD modelling results as well, which eventually affects the forecasting accuracy. From non-radial propagation of CMEs in the lower corona \citep{2024Valentino} to human-in-the-loop errors \citep{2023Verbeke}, the estimation of geometrical parameters close to the Sun translates into big uncertainties in predictions at 1~au \citep{2012Kilpua,2015Mays,2018Wold,2021Riley,2024Kay}. Current forecasting accuracy of the CME arrival time at Earth, regardless of the method or model used, ranges around $\pm 10$~hours \citep[e.g., ][]{2018Wold,2019Vourlidas,2024Rodriguez}.
    
Novel 3D visualisations approach in the studies of the ambient solar wind and its origins \citep[e.g., ][]{2021Samara}, show that the solar wind streams' sources can be relatively compact when seen closer to the inner boundary of the model. Nevertheless, as the solar wind expands into the heliospheric domain, solar wind streams' shape evolves differently depending on the latitude and the ambient conditions. In this work, we take a similar 3D approach for inspecting both the ambient solar wind and embedded CMEs. We focus on the effects of the interaction of the solar wind streams with the CMEs and CME-driven shocks. The results shown in this work were generated using EUHFORIA and the adapted cone model \citep{2018Scolini}. In order to better understand how the deformations of the CME-driven shock, due to the interaction with the ambient solar wind, are formed, we positioned multiple virtual spacecraft in the simulation domain to follow the evolution of the CMEs three-dimensional structure \citep{2018Lugaz}. This approach was previously used by \citet{2021Scolini} and follows the methodology used in \citet{2024Valentino}. A somewhat similar approach, with the CME-solar wind interaction addressed using a 3D MHD model, was employed by \citet{2024Mayank}. The authors employed SWASTi-CME model \citep{2024Mayank} and two different CME models (modified cone model and FRi3D).

This paper focuses on the non-radially propagating CME events of December 7, 2020, and October 28, 2021, previously studied by \citet{2024Valentino}. As shown there, in situ measurements show signatures of both events at Earth, although the CME's main direction of propagation was not along the Sun-Earth line. In this work, we analyse the role of the background solar wind in shaping the CME and CME-driven shock expansion and impacting their propagation. To achieve this, we simulate both events using the 3D MHD model EUHFORIA \citep{Pomoell2018, 2020Poedts} and the cone CME model \citep{2004Xie,2018Scolini}. We also utilised a shock tracing method, where the discontinuity generated by the cone CME is located in the 3D EUHFORIA domain. 

We present the models and methods used in this study in Section~\ref{sect:EUHFORIA_default}. A summary of the studied events and a detailed motivation for this multidimensional approach is provided in Section~\ref{Sect:Event_Description}, while the modelling results in 3D, 2D, and 1D can be found in Section~\ref{sect:modelling_3D_2D_1D}. By using the full 3D MHD output, we can locate where the CME drives a shock and calculate the gas compression ratio, which determines the strength of the shock (Section~\ref{sect:CME_driven_shock_structure}). Using these results, we traced the CME-driven shock evolution in time as presented in Section~\ref{sect:evolution_CME_Structure}. Finally, we discuss the results of this study and present the summary and conclusions in Section~\ref{sect:Summary_and_conclusions}.



\section{Models and methods}
\label{sect:EUHFORIA_default}
    This study focuses on two halo CMEs observed on December 7, 2020, and October 28, 2021. Both CMEs had non-radial propagation directions in the low corona, and their main propagation direction in the inner heliosphere was not fully aligned with the Sun-Earth line. Detailed analysis of these two events is presented in \citet{2024Valentino}. The authors combined remote sensing and in situ observations to constrain CME input parameters for modelling with 3D MHD solar wind and CMEs model EUHFORIA \citep{Pomoell2018, 2020Poedts}. The aim of the study was to improve our understanding of how the choice of initial CME parameters affects simulation accuracy at Earth distances. However, \citet{2024Valentino} did not address the possible changes of the shape of the CME due to its interaction with the background solar wind, as this is difficult to quantify using time-series analysis alone. 
    In this study, we inspect the impact of the background solar wind on the 3D structure of the CME and the CME-driven shock by using the full 3D MHD simulation output of the state-of-the-art MHD code EUHFORIA.

    \subsection{Numerical set-up of the EUHFORIA model}

    The steady-state 3D MHD code EUHFORIA model, in its default setup \citep{Pomoell2018, 2020Poedts}, models the solar wind and CMEs in the inner heliosphere up to the distances of 2~au. This flexible model consists of 3 main parts: the coronal part, the heliospheric part, and the CME-insertion part. The coronal part uses as input global photospheric magnetic field maps that have been constructed based on synoptic magnetograms from observations of the full solar surface (360$^{\circ}$). Reconstruction of the corona is based on the potential-field source surface extrapolation, the Schatten Current Sheet, and the Wang-Sheeley-Arge models \citep{PFSS1969,Schatten1969,Wang1990,2000Arge}. The GONG \citep[Ground Oscillation Network Group; ][]{1996Harvey,2018Hill} full-disk solar magnetic field maps are the ones that are most often employed as an input to EUHFORIA but other types of magnetograms, like the GONG-ADAPT \citep[GONG Air Force Data Assimilative Photospheric Flux Transport; ][]{2000Worden,2009Arge,2015Hickmann} or the HMI \citep[Heliospheric and Magnetic Imager; ][]{2012Scherrer} on board the SDO \citep[Solar Dynamics Observatory; ][]{2012Pesnell} can also be chosen. The output of the coronal part of EUHFORIA, so-called EUHFORIA's inner boundary conditions, are the plasma and the magnetic field conditions that are further used as an input for the heliospheric part, which solves the ideal MHD equations augmented with gravity.

    The CMEs are inserted in EUHFORIA at the inner boundary of the model's heliospheric part, i.e., at 0.1~au. A number of different CME models can be used in EUHFORIA, such as the cone model \citep{2004Xie}, linear force-free spheromak \citep{2019Verbeke}, Flux Rope in 3D \citep[FRi3D; ][]{2016Isavnin,2022Maharana}, and horse-shoe model \citep{2024Maharana}. All of them need input parameters for the position and the structural characteristics of the CMEs, and spheromak, FRi3D, and horse-shoe models also need information about CMEs' magnetic properties.
    
    EUHFORIA simulations were performed using a $1^{\circ}$ angular resolution, 308 cells in the radial direction up to 1.2~au, with a time resolution of 15 minutes. In this study, we employed a smaller simulation domain than that used in \citet{2024Valentino}, opting to prioritise a higher temporal cadence of the model output. While the radial resolution remains comparable to the previous work, the high temporal and angular resolutions allowed for a more detailed mapping of the 3D structures of the solar wind and CME.  

    \subsection{CME cone model \& terminology}
    \label{sect:Cone_description}
 
    During the last few decades, the most frequently employed CME model was the cone model \citep{2004Xie, 2013Millward,2002Zhao}. In order to better understand and put in context the results of several decades of modelling with the cone model, we employed the cone CME model also in this work. The cone CME is defined as a hydrodynamic, uniformly filled cloud with the pressure, speed, temperature and density being constant within the CME during its insertion. The version of the cone model employed in this study consists of two parts \citep{2018Scolini}. Its lower part is a circular cone that has its vertex, i.e., its narrowest point, in the centre of the Sun. Its upper part, which constitutes the front of the Cone, is of spheroidal shape, and it is attached to the base of the lower part. This CME shape is similar to the "full ice cream cone model" employed in some other works \citep[see e.g., ][]{2005Xue, 2009Gopalswamy}. Since the cone model does not consider the magnetic structure of the CME, it is not appropriate for studies of the CME's magnetic properties or its geomagnetic impact. The focus of this study is on the interaction of the background solar wind and CMEs, which does not necessarily require the magnetic characteristics of the CMEs. The case of interaction of the magnetic CMEs with the ambient solar wind will be addressed in the follow-up publication. In this study, by using the cone model we are trying to advance our knowledge of CME-solar wind interactions, and put this knowledge in the context of a few decades of cone CME modelling results. In this way we also make more comprehensible the often-found large uncertainties of the CME and CME-driven shock wave arrival times, crucial in the space weather forecasting.

    The input parameters of the cone model that define its position and size at the EUHFORIA's inner boundary are the longitude and latitude, while the half angular width represents the half angle of the CME cone. The herein employed values were obtained by using the Graduated Cylindrical Shell fitting technique \citep[GCS; ][]{2006Thernisien, 2009Thernisien, 2011Thernisien} and they are the same as one in \cite{2024Valentino} which provided the most accurate arrival time at 1~au. The details about the process for obtaining the insertion time of the CME at the inner boundary and its speed can be found in \cite{2024Valentino}. For the temperature and the density, we used the values of $T_{CME}$~=~0.8~MK and $\rho_{CME}$~=~10$^{-18}$~kg~cm$^{-3}$ respectively, proposed in \cite{Pomoell2018}. We summarise in Table~\ref{tab:Run_Parameters} the information about the selected magnetograms and CME input parameters. 

\begin{table}[h!]
        \centering
        \begin{tabular}{|c|c|c|}
        \hline
            \textbf{Parameters} & \textbf{Event~1} & \textbf{Event~2}\\
        \hline
        \hline
            Date of magnetogram & 2020/12/07 & 2021/10/27\\
            (YYYY/MM/DD) & & \\
        \hline
            Time of magnetogram & 10:00 & 19:04\\
            (HH:MM UT) & & \\
        \hline
        \hline
             Date of CME& &\\
             (YYYY/MM/DD) & 2020/12/07  & 2021/10/28\\
        \hline
             Time of insertion & &\\
             (HH:MM UT)& 19:32 & 19:09\\
         \hline
             Latitude~($^\circ$)& -24.6 &-44.0 \\
         \hline
             Longitude~($^\circ$)& 18.3 & 0.0 \\
         \hline
             Half-Width~($^\circ$)& 29.4 & 49.0\\
         \hline
             Speed~($\mathrm{km~s}^{-1}$)& 1236 & 1005\\
         \hline
             Mass~density~($\mathrm{kg~m}^{-3}$)& 1$\times$10$^{-18}$  & 1$\times$10$^{-18}$ \\
         \hline
             Temperature~(K)& 8$\times$10$^{5}$  & 8$\times$10$^{5}$\\
         \hline
        \end{tabular}
        \caption{Information about the magnetogram and the CME parameters for the two selected events, as obtained in \citet{2024Valentino}.}
        \label{tab:Run_Parameters}
    \end{table}

    \subsection{Shock tracing method}
        \label{sect:Shock_tracing_method}

    Shock waves are discontinuities, characterised by the sudden and simultaneous increase of the plasma characteristics, plasma density, temperature, pressure, speed, and magnetic field. The Rankine-Hugoniot relations connect the upstream and the downstream plasma properties and place a limit on the possible values of these quantities across the shock \citep{2023Gedalin}. In particular, the CME-driven shocks have to satisfy these relations. To be able to follow these structures described in Sect.~\ref{sect:CME_driven_shock_structure} and \ref{sect:evolution_CME_Structure}, we isolated them and analysed in greater detail using the automatic shock-tracing algorithm presented in Appendix~A of \citet{2022Wijsen}. This method isolates the CME-driven shock and removes any shock associated with corotating interacting regions (CIRs) by subtracting the unperturbed simulated background solar wind from the simulated solar wind with the CME of interest included. This is only possible because EUHFORIA's solar wind is steady state in a frame corotating with the Sun. This subtraction can be performed on any model variable or derived quantity. In particular, the entropy ($S\equiv P/\rho^{\gamma}$, with $P$ the thermal pressure, $\rho$ the density and $\gamma$ the adiabatic index) is a quantity of interest for the study. After doing the subtraction, relative quantities, such as the relative velocity ($V_{rel}$) and the relative entropy ($S_{rel}$), are calculated. With the term ``relative'' we refer to the solar wind properties (e.g. density, temperature, or heat flux) calculated as the relative change of a quiet background plasma characteristics. To reconstruct the full surface that represents the CME-driven shock, the algorithm finds along the main direction of propagation of the CME the furthest cell, where $V_{rel}>\epsilon_{V}$ and $S_{rel}>\epsilon_{S}$\footnote{$\epsilon_{V}$ and $\epsilon_{S}$ are predefined positive values that are required due to the numerical accuracy of the simulations.}. Additionally, the divergence of the solar wind ($\nabla \cdot V$) is required to be negative at this position to ensure that a compression region is located. Then, the shock surface is located by calculating the isosurface of $S_{rel} = \epsilon_{S}$ that includes this point along the main propagation direction using a marching cubes algorithm included in the open-source Python library PyVista \citep{Sullivan2019}. This method has been employed to capture the complexities of CME-driven shock structures in studies of solar energetic particle (SEP) events \citep{2022Wijsen,2023Wijsen,2023Niemela,2024Niemela}. 
    
    This method allows us to calculate the gas compression ratio ($r_{g}$) between the downstream and the upstream plasma density and extracting the CME-driven shock surface from the full 3D EUHFORIA simulations. In regions where $r_{g} \leq 1$, the speed is lower than the fast magnetosonic speed, and therefore not part of the CME-driven shock wave. On the other hand, regions when $r_{g} > 1$, are considered as part of the CME-driven shock wave. We employed this method on both simulations, and the results are showcased and analysed in Sect.~\ref{sect:CME_driven_shock_structure}. 

\section{Events description and motivation for the study} 
\label{Sect:Event_Description}
    In this study, we focused on two CME/flare events observed on December 7, 2020 and on October 28, 2021, referred further as Event~1 and Event~2, respectively.
    Details about the events and modelling with several different modelling setups were presented in \cite{2024Valentino}. Here, we provide a short description of the events and the EUHFORIA modelling results. It is important to note that, although both studied events had a main propagation direction southward from the Sun-Earth line, Event~2 had somewhat more complex background solar wind. 
    
    For modelling both solar wind and CMEs with EUHFORIA, we employed the input parameters that provided the most accurate CME arrival time in \citet{2024Valentino}. 
    The general information on the modelling set-up, which is important for this study, is listed in Sect.~\ref{sect:EUHFORIA_default}. As this study aims also to provide knowledge on the operational forecasting aspects of EUHFORIA, we employ its so-called default set-up \citep{Pomoell2018,2019Hinterreiter}, and we do not tune the coronal model to these specific events.

    \subsection{Event 1: December~7~2020}
    \label{sect:EUHFORIA_default_Dec2020}

    The CME on December 7, 2020, was first observed in the SOHO/LASCO C2 coronagraph images \citep[LASCO - Large Angle and Spectroscopic COronograph and C2 on board SOHO - SOlar and Heliospheric Observatory; ][respectively]{1995Brueckner,1995Domingo} at 16:24~UT.
    The STEREO-A COR~2 coronagraph \citep[Solar TErrestrial RElations Observatory Ahead; ][]{2008Howard,2008Kaiser} observed the CME somewhat earlier, starting from 15:54~UT. The long-duration C7.4 GOES X-ray flare \citep[Geostationary Operational Environmental Satellite; ][]{1994Garcia} associated with the studied CME originated from the NOAA AR~12790 (S23W14).
    Rather small coronal dimmings associated with the CME lift-off were observed by SDO/AIA \citep[the Atmospheric Imaging Assembly on board Solar Dynamics Observatory; ][]{2012Lemen, 2012Pesnell} at 304 and 193~\AA, while the eruption of the related filament was observed at 211~\AA. The type~II radio burst, which indicates the existence of the CME-driven shock, was observed by both STEREO~A/WAVES \citep{2008Bougeret} and WIND/WAVES \citep{1995Bougeret} instruments. The CME-driven shock was observed in the ACE \citep[Advanced Composition Explorer; ][]{1998Stone} and WIND \citep{1995Harten} in situ data at Earth as a flank encounter at 01:30~UT on December 10, 2020. The observed values match all the criteria according to the database of heliospheric shock waves \citep{2015Kilpua}, and it is also listed with 99~\% confidence level in the \href{https://space.umd.edu/pm/figs/figs.html}{SOHO/Celias} in-situ shock database. 

    \subsection{Event 2: October~28~2021}
    \label{sect:EUHFORIA_default_Oct2021}
    The CME on October 28, 2021, first appeared in the field of view of SOHO/LASCO C2 coronagraph at around 15:48~UT and at around 15:53~UT in the field of view of STEREO-A~COR~2 coronagraph. The associated GOES flare was more energetic for this event than in the case of Event~1. The X1.0 class flare, which lasted several hours, originated from the NOAA AR~12887 (S26W07). More about the associated particle event can be found in \cite{2022Klein}. EUV observations by SDO/AIA at 193 and 211\,\AA\, show large coronal dimming at the north of the AR and an EIT wave that propagated mostly towards the north from the eruption site. The type II radio burst started in the metric wavelength range at around 15:30~UT and continued at larger wavelengths, i.e., in WIND/WAVES and STEREO~A/WAVES observations, with a duration of about 20 to 30 minutes, respectively. The radio observations therefore show the shock wave existence from the low corona up to interplanetary space. The arrival of the CME-driven shock at Earth was observed in the in situ data at around 09:30~UT on October 31, 2021.

    \subsection{Motivation for the multidimensional approach} 
    \label{sect:motivation}

\begin{figure*}[h!]
        \includegraphics[width=0.98\textwidth]{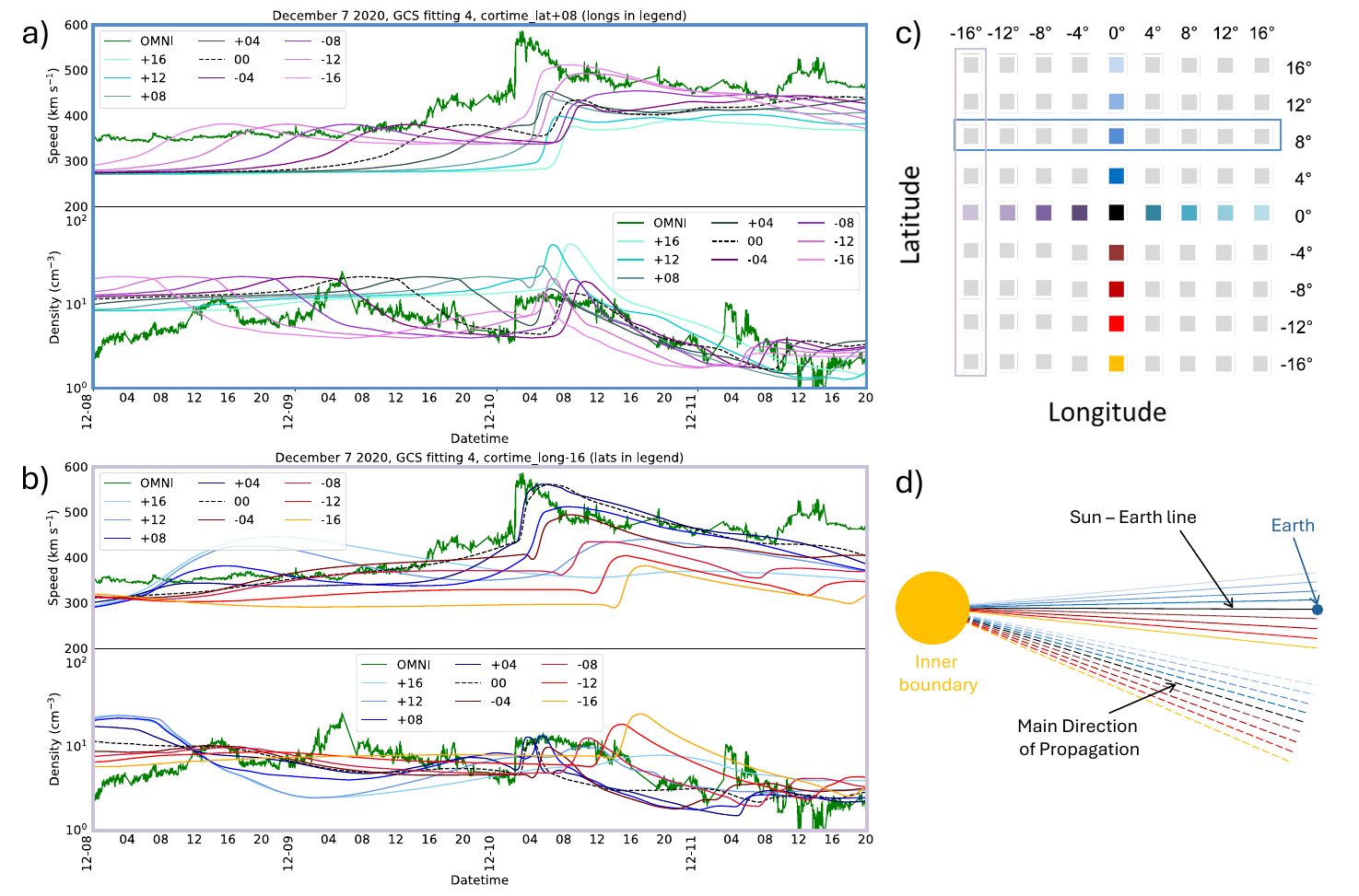}
        \caption{Time series at positions of 9 virtual spacecraft above and below the Sun-Earth line with $\pm$16$^{\circ}$ longitudes are shown in panels a and b, respectively. These plots demonstrate how difficult it is to understand the overall shape of the modelled CME front when taking into consideration only 1D time profiles. In panel c, we show the grid of virtual spacecraft that we used to plot different 1D plots. Sketch on the panel d shows the two directions in which we positioned the grids.}
        \label{Fig:motivation_plots}
    \end{figure*}

     Forecasting CMEs is very important in both the space weather studies and operations. Interaction of CMEs with the ambient solar wind can result in their structuring and potentially different arrival times at Earth. Recent observations show that CME and CME-driven shock time profiles can sometimes be similar and sometimes very different, within small spatial and temporal scales \citep[][respectively]{2021Davies,2024Palmerio}. In order to inspect the 3D structure of the CMEs, and thus also the effect of their interaction with the solar wind, we employ modelling with EUHFORIA. We first use the 1D time series, which are traditionally used as they can be directly compared with observations. We plotted results of the simulations in 1D for Event~1 (see also \citealt{2024Valentino}), using the grid of virtual spacecraft with 4$^{\circ}$ spacing that span from -16$^{\circ}$ to +16$^{\circ}$ in both longitudinal and latitudinal directions (see Fig.~\ref{Fig:motivation_plots}c). While \cite{2024Valentino} used only one set of virtual spacecraft at different latitudes with the Earth as the central position, we also inspected modelling results at virtual spacecraft with different longitudes as the central positions. In Fig.~\ref{Fig:motivation_plots}a and b, we show plots for Event~1, the case of latitude at +8$^{\circ}$ and for longitude at -16$^{\circ}$, respectively. The modelling time series show very complex profiles, although central positions are not very widely spaced, not allowing us to understand the exact shape and structure of the CME in the 3D space, even if we consider the smallest angular separation (4$^{\circ}$) between different virtual spacecraft. Based on these plots, we concluded that 1D time series are not sufficient for understanding the complex impact of the CME-solar wind interactions on the structure of the CME. To better understand and try to quantify the spatial extent of the CME and CME-driven shock wave deformations found in the 1D time profiles, we inspect modelling results in the 1D, 2D, and 3D domain. 

    \section{Modelling CMEs in the 3D, 2D and 1D domain} 
    \label{sect:modelling_3D_2D_1D}

    In this section, we present the simulation results of both studied events and discuss the modelled structures of the CME front. We present them in 3D and 2D plots to better visualise and understand the interaction between the solar wind and the CME. To try to quantify the interaction effect and to have modelling results comparable with in situ observations, we also show 1D plots. Since modelling is done in the 3D domain, we are able to extract the 1D time series at any needed time-space point, which provides us with a comprehensive view of the structured CME. 

    To meticulously follow CME evolution for both studied events, we placed the 4$^{\circ}$ grids (Fig.~\ref{Fig:motivation_plots}c) along the two propagation directions, with virtual spacecraft positioned at different radial positions along these directions. Figure~\ref{Fig:motivation_plots}d) shows a sketch of the two different sets of directions used for the modelling. CME propagation directions along the Sun-Earth line are presented with the solid lines, and the one along the main propagation direction (MDP) with the dashed lines.

    \begin{figure*}[h!]
            \includegraphics[width=0.98\textwidth]{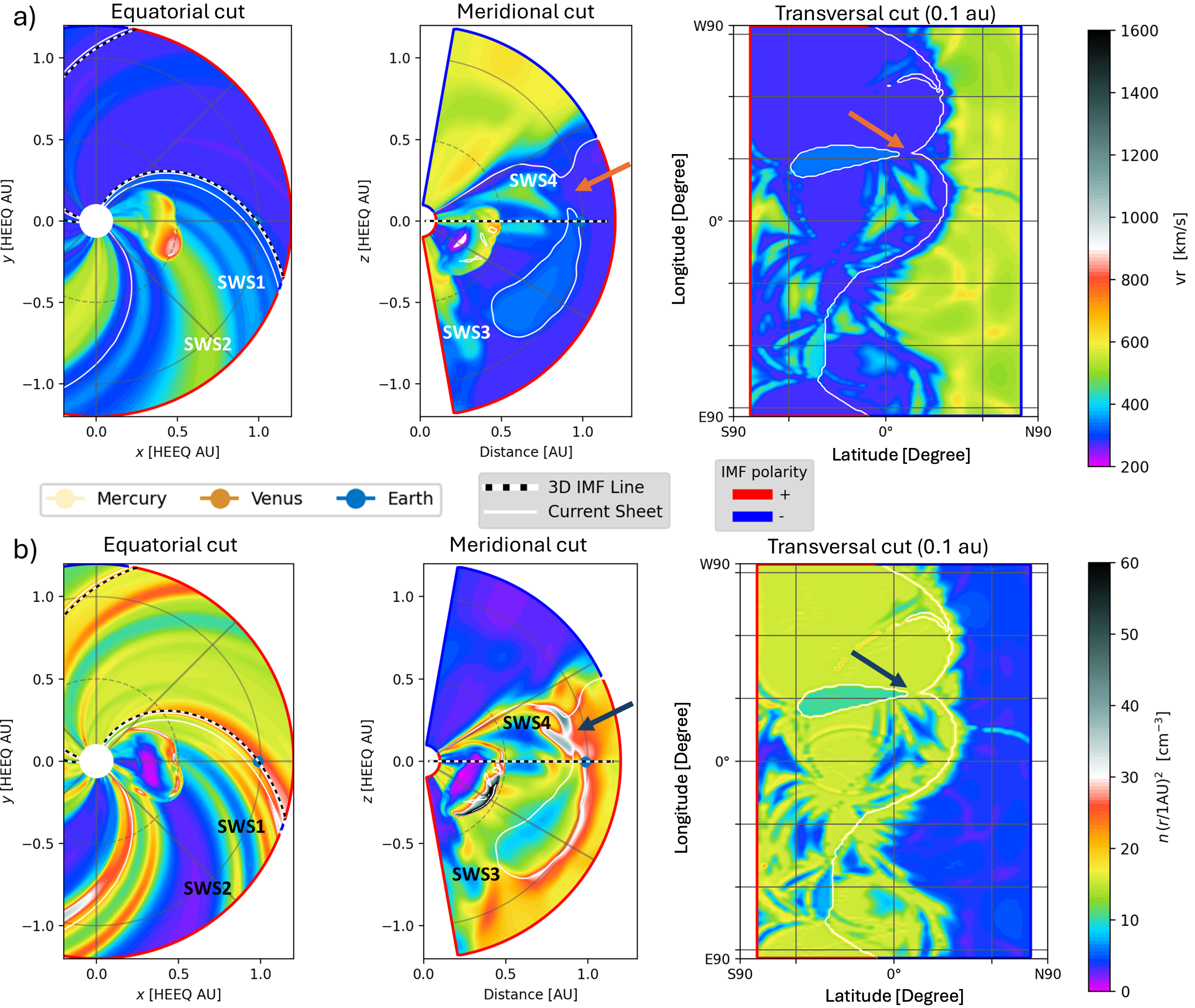}
            \caption{EUHFORIA simulation output for Event~1 on December~8,~2020 at 14:48~UT. Panel a shows equatorial, meridional (at Earth’s longitude), and transversal (at 0.1 au) cuts of the radial solar-wind speed. Panel b shows the corresponding cuts for the scaled plasma density. The thin white line corresponds to the position of the heliospheric current sheet (HCS), and the dashed line shows the magnetic connection to Mercury, Venus and Earth, calculated with the Parker spiral and using the local speed at these locations. The approximate position of the Solar Wind Streams (SWS) are annotated on both panels, together with an arrow pointing at the narrowing of the HCS, a key characteristic of this Event.}
            \label{Fig:Default_EUHFORIA_Dec2020}
        \end{figure*}

    \subsection{Modelling results in the 2D and 3D domains} 
    \label{subsect:modelling_3D_2D_1D}

        \subsubsection{Event~1: December~7~2020}
        \label{subsubsect:4.1_Event1}
     
\begin{figure*}[h!]
        \includegraphics[width=0.95\textwidth]{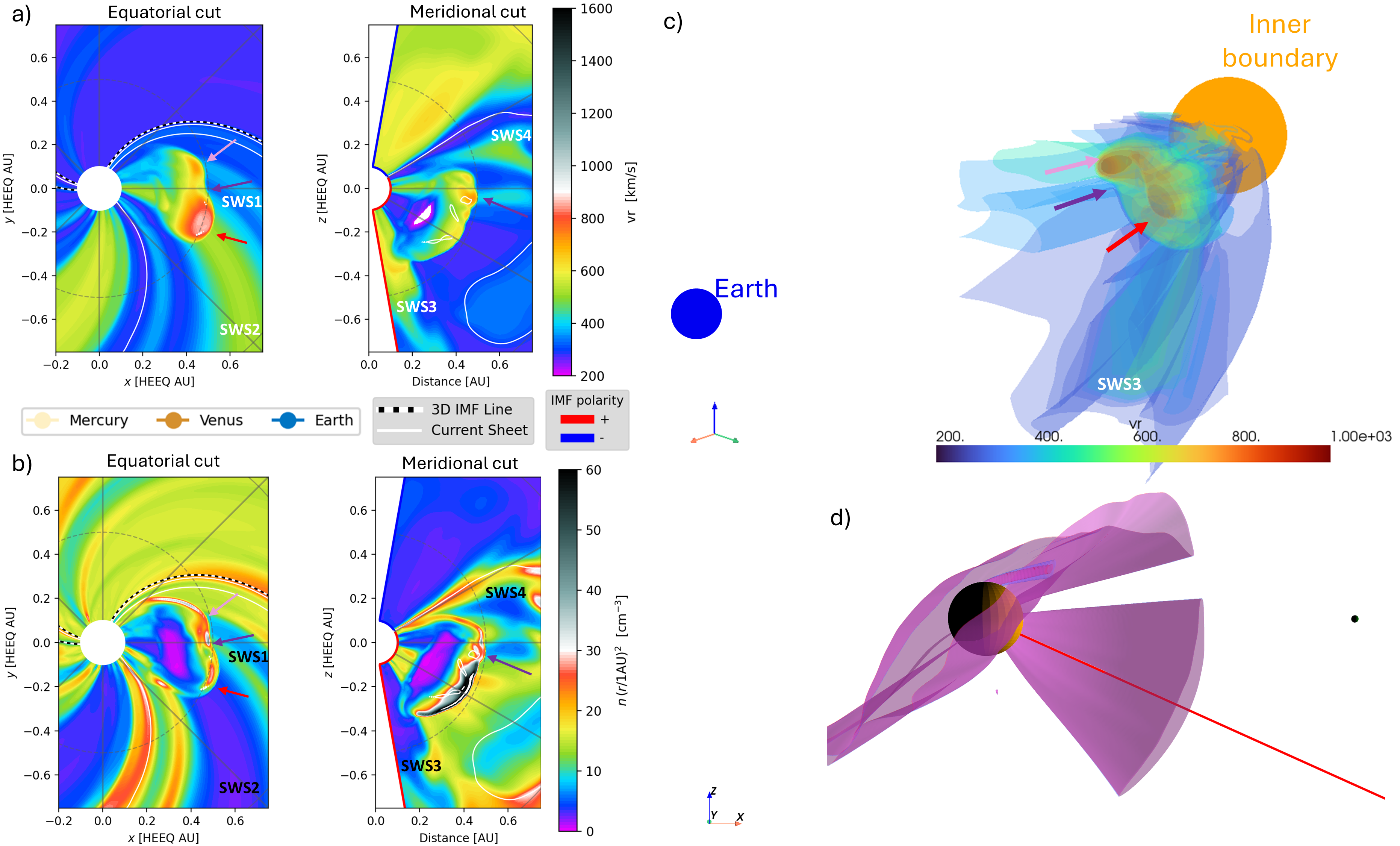}
        \caption{2D visualisation of EUHFORIA simulation results at 14:48~UT, on December~8,~2020 for solar wind speed and scaled density in a case of Event~1 (panels a and b, respectively). The zoomed figure shows in detail the modelled CME and allows better comparison with the 3D presentation. The 3D visualisation of EUHFORIA's simulation of solar wind speed and the structure of the heliospheric current sheet (panels c and d, respectively). The presented HCS structure is obtained as a result of the solar wind modelling without inserted CME, so its shape does not reflect the CME-HCS interaction. Coloured arrows point at different features that can be seen in both the 2D and 3D plots.}
        \label{Fig:2020Event_with_arrows}
        \end{figure*}

        The 2D EUHFORIA simulation results for Event~1 are presented in Fig.~\ref{Fig:Default_EUHFORIA_Dec2020}, with a zoomed-in view in panels a and b of Fig.~\ref{Fig:2020Event_with_arrows}. Panel a of both Figs. shows the equatorial, meridional, and transversal plane (only in Fig.~\ref{Fig:Default_EUHFORIA_Dec2020}) cuts for radial speed $v_r$, while the lower panels show the same cuts but for scaled number density $n$, both at distances $0.1-1.2$~au. The equatorial and meridional plane cuts include Earth, while the transversal cut is made at 0.1~au. The modelled background solar wind in which Event~1 propagates is relatively simple with only a few distinctive solar wind streams (SWS). We employ the term SWSs and not high speed streams (HSS) because it is generally considered that the HSS have speeds at and above 500~km~s$^{-1}$. 
        EUHFORIA, as well as other MHD models with a similar set-up, tends to underestimate the fast solar wind speed and/or the latitudinal and longitudinal extent of the wind flows \citep{2019Hinterreiter,2019Asvestari}, although the complexity of the modelled solar wind is generally still captured. The fast wind speed underestimation is closely linked to the area and morphology of the coronal holes sourcing the fast wind \citep{2020Hofmeister,2022Hofmeister,2022Samara}. Due to this modelling characteristic, we employ herein the more general term solar wind streams i.e., SWS, instead of the term HSS. Figure~\ref{Fig:Default_EUHFORIA_Dec2020} shows four distinctive solar wind flows, SWS~1, SWS~2, SWS~3, and SWS~4 modelled by EUHFORIA, in the propagation direction of the CME. We note that not all of the SWSs are present in the different plane cuts. At 1~au, the solar wind stream SWS~1 has a scaled number density close to 25~cm$^{-3}$ and speeds in the order of 350~km~s$^{-1}$ while the solar wind stream SWS~2 has scaled number densities close to 5~cm$^{-3}$ and speeds in the order of 500~km~s$^{-1}$. Close to 0.75~au, and only present in the meridional cut, SWS~3 has a scaled number density close to 5~cm$^{-3}$ and speeds in the order of 500~km~s$^{-1}$, while SWS~4 shows scaled number densities close to 5~cm$^{-3}$, and speeds in the order of 550~km~s$^{-1}$.

        The HCS (Heliospheric Current Sheet), shown as a thin dashed white line (Fig.~\ref{Fig:Default_EUHFORIA_Dec2020} and Fig.~\ref{Fig:2020Event_with_arrows} panels a and b), has a relatively simple structure, as seen in the 2D presentation. At Earth's location, the HCS is inclined approximately 45$^{\circ}$. The modelled HCS shows a closed structure, clearly seen in the transversal and in the meridional cut. Figure~\ref{Fig:2020Event_with_arrows}d shows even more comprehensively the particular closed structure of the HCS in the 3D presentation. Two distinguishable parts of the HCS are a typical ballerina-skirt-like structure and a tubular closed structure. The closed structure, seen both in 2D and 3D presentations, is probably part of a strongly wrapped HCS, which is in reality connected to the ballerina-skirt part of the HCS in the thin region, which is not modelled correctly by EUHFORIA. The place where we expect the closed magnetic structure to be connected to the HCS is in Fig.~\ref{Fig:Default_EUHFORIA_Dec2020} marked with the orange arrows. The reason for such particular modelling results is the shape and structure of a strongly warped HCS with the thin corridor, which is not properly modelled (see Fig.\ref{Fig:Default_EUHFORIA_Dec2020}) due to numerical limitations, and processing of the input magnetogram. Although \cite{2014Wang} showed that structures like that are possible during solar maximum, we don't consider that here this is the case. The events in the present study took place during the ascending phase of the 25th cycle, when the global magnetic field did not show much complexity. Due to its main direction of propagation, the CME only crosses the tubular-like closed surface and not the ballerina part of the HCS. 

        As illustrated in Tab.~\ref{tab:Run_Parameters} and in the meridional plane cut in Fig.~\ref{Fig:Default_EUHFORIA_Dec2020}, CME of Event~1 propagates strongly southward from the Sun-Earth line. The equatorial plane cut in Figs.~\ref{Fig:Default_EUHFORIA_Dec2020} and Figure~\ref{Fig:2020Event_with_arrows} shows that the CME is well visible in the first and fourth quadrants, indicating that CME has a significant component in the eastward direction from the Sun-Earth line. 

        Panels~a and b of Fig.~\ref{Fig:2020Event_with_arrows} show velocity and scaled number density, respectively, in the 2D EUHFORIA simulation results for Event~1. Similarly to Fig~\ref{Fig:Default_EUHFORIA_Dec2020}, we focus on the equatorial and meridional plane cuts, but the modelling domain is presented up to distances of 0.7~au to better visualise the details of the CME and ambient solar wind structure. Panel~c shows the 3D isosurfaces for the radial solar wind speed, and panel~d shows only the 3D structure of the HCS. We note that although the modelling was done for the full 3D space, the plotted area in panel c is only in the neighbourhood of the CME, in order to more clearly show the CME and solar wind interaction. The colourful arrows in all panels point to the solar wind flows and different parts of the CME that are of interest for the study. The orange sphere presents the inner boundary of the EUHFORIA's modelling domain.  The blue circle in panels c and d represents the Earth's position (its size is not to scale), indicating that the CME is Earth-directed. 
        
            
        The CME front modelled by EUHFORIA is seen in Fig.~\ref{Fig:2020Event_with_arrows}, panels a) and b), as a sharp, wavy, croissant-like structure in the ambient solar wind. Different parts of the CME front propagate in the different solar wind regimes, which results in the structuring, clearly seen in the 2D speed and density plots. The distortion is visible in both considered cuts. A bump of the CME front in the vicinity of the ecliptic plane is most prominent in the equatorial cut and the 3D isosurfaces (Fig.~\ref{Fig:2020Event_with_arrows}a and c). The knob is clearly seen in the indent between the two northern knobs of the CME front, marked with the pale-pink and red arrows. These two knobs are fully emerged in SWS~1 and SWS~2, and the interaction between these parts of the CME and the background solar wind is more pronounced than in the CME flank regions. The concave part of the front, marked with the violet arrow, exhibits a curvature change that reflects this same interaction with a different type of solar wind at that location. In contrast, the southward-directed SWS~3 is not seen in the equatorial cut, due to its high latitude, but it is well distinguishable in the meridional plane cut and the 3D solar wind isosurface. SWS~3 only interacts with the outer part of the CME flank, and, since the orientation of the flank region and of SWS~3 are quite similar, this interaction does not significantly influence the CME shape and expansion.
        
        \subsubsection{Event~2: October~28~2021}

\begin{figure*}[h!]
                \includegraphics[width=0.98\textwidth]{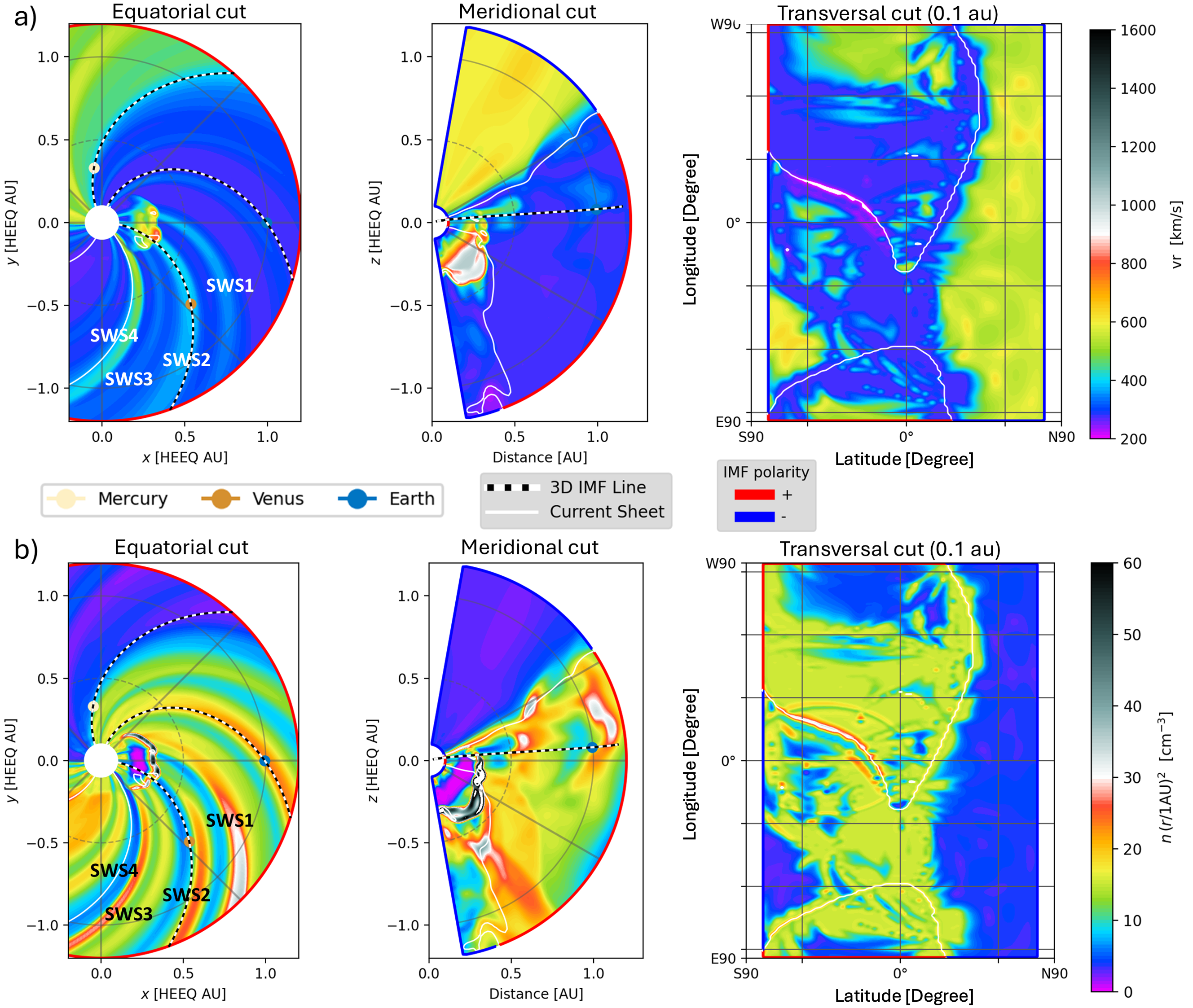}
                \caption{EUHFORIA simulation output for Event~2 on October~29,~2021 at 11:13~UT. Panel a shows equatorial, meridional (at Earth’s longitude), and transversal (at 0.1 au) cuts of the radial solar-wind speed. Panel b shows the corresponding cuts for the scaled plasma density. The thin white line corresponds to the position of the heliospheric current sheet (HCS), and the dashed line shows the magnetic connection to Mercury, Venus and Earth, calculated with the Parker spiral and using the local speed at these locations. The approximate position of the Solar Wind Streams (SWS) are annotated on both panels.}
                \label{Fig:Default_EUHFORIA_Oct2021}
            \end{figure*}

        Similarly to Sect.~\ref{subsubsect:4.1_Event1} about Event~1, we first present the modelling results in the 2D space (Fig.~\ref{Fig:Default_EUHFORIA_Oct2021}). EUHFORIA's modelling results show quite variable solar wind with four solar wind streams, SWS1, SWS2, SWS3 and SWS4 in the direction of the Sun-Earth line, with their major part being below the ecliptic plane. The naming of those streams has the same motivation as for Event~1. At 1~au, all four streams of Event~2 have comparable properties, with speeds of about 400--450~km~s$^{-1}$ and scaled number densities of about 5--10~cm$^{-3}$.
        
\begin{figure*}[h!]
                    \includegraphics[width=0.98\textwidth]{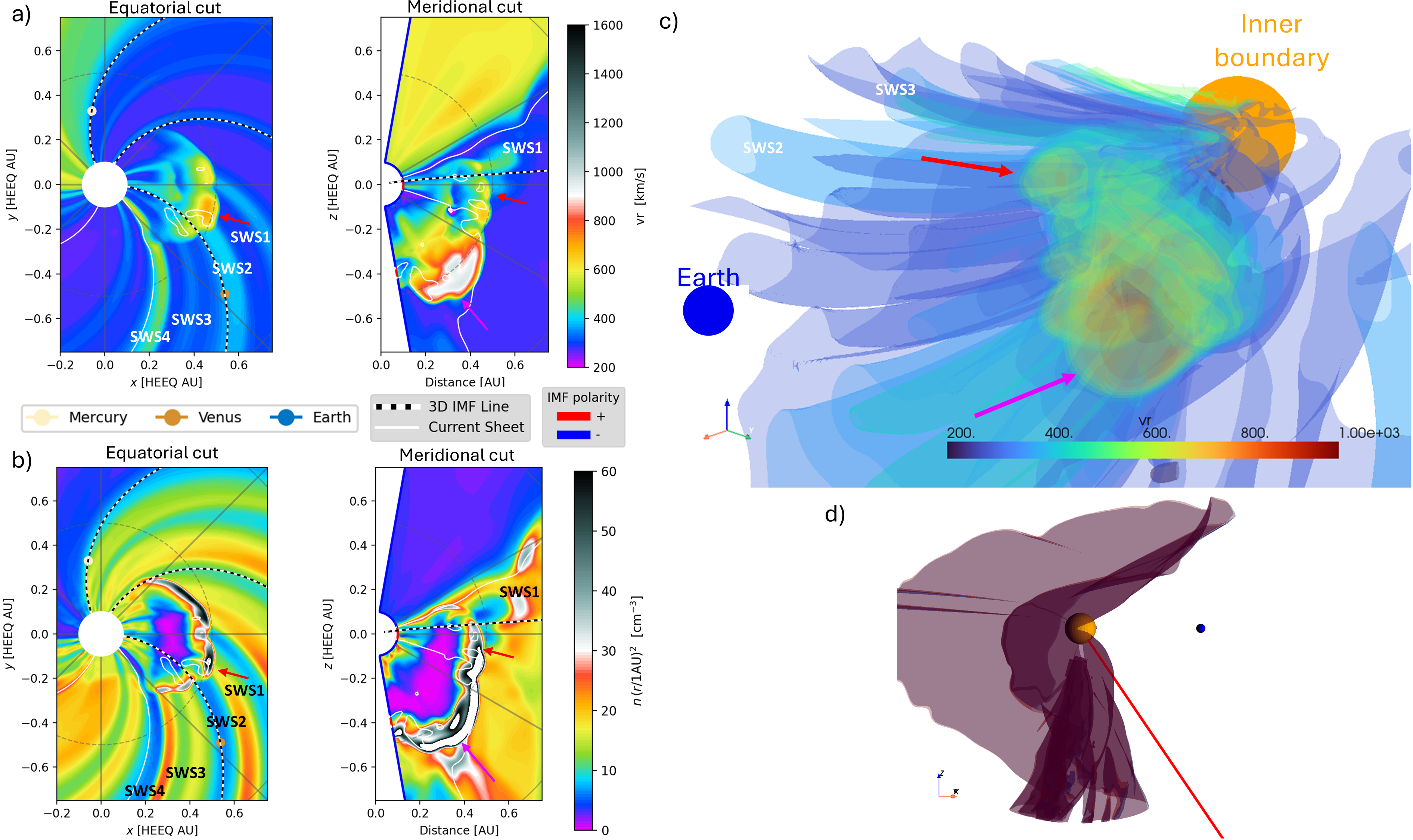}
                    \caption{2D visualisation of EUHFORIA simulation results for October~29,~2021 at 20:13~UT for solar wind speed (panel a) and scaled density (panel b) for Event~2. 3D visualisation of EUHFORIA simulation of solar wind speed (panel c) and the structure of the heliospheric current sheet (panel d). The last panel displays the simulation results for a snapshot taken before the CME injection; therefore, the HCS structure is not influenced by the CME. Coloured arrows point at different features that can be seen in the 2D and 3D plots.}
                    \label{Fig:2021Event_with_arrows}
                \end{figure*}

        The HCS (Fig~\ref{Fig:2021Event_with_arrows}d) has a somewhat more complex configuration than in Event~1. It is visible as inclined about $ \pm45^{\circ}$ in the meridional plane, and situated along the CME's propagation direction. The CME crosses the HCS with its southern part, which is visible in Fig.~\ref{Fig:2021Event_with_arrows}d, where the line that marks the CME's main propagation direction intersects with the HCS surface at least two times. 
        As shown in Table~\ref{tab:Run_Parameters} and in the meridional plots of Fig.~\ref{Fig:Default_EUHFORIA_Oct2021},  Event~2 was propagating strongly southward, with respect to the Sun-Earth line. Its main direction of propagation was about 44$^{\circ}$ off the equatorial plane. In the equatorial plots, the CME appears as centred at the Sun-Earth line. However, the equatorial cut crosses only the northern CME flank, as can be seen from the meridional plane cut (Fig.~\ref{Fig:Default_EUHFORIA_Oct2021}). The strong structuring of the background solar wind along the main CME propagation path resulted in a very structured CME front, as visible in the zoomed 2D plots (Fig.~\ref{Fig:2021Event_with_arrows}\footnote{It is important to note that Fig.~\ref{Fig:Default_EUHFORIA_Oct2021} and \ref{Fig:2021Event_with_arrows} correspond to different times in the simulation.}). Two main CME lobules can be observed in the ecliptic plane cut, meridional plane cut, and in the 3D presentation, marked with the magenta and red arrows. Since the CME structure has a very complex shape, structures of a generally different shape are visible in the ecliptic and meridional plane cuts, e.g., the well-defined CME lobe marked with the magenta arrow is only visible in the meridional cut and in the 3D presentation. The same holds for different solar wind flows, as can be seen when comparing the 3D presentation with the 2D plots. The solar wind flow that has the largest influence on the deformation of the CME is SWS~2, which propagates mostly southward from the ecliptic plane, i.e., in the same direction as the studied CME. We note that due to projection effects and apparent overlap with SWS~1, the SWS~2 does not seem very prominent in the 3D presentation.

    \subsection{Modelling in 1D: time profiles} 
    \label{subsect:1D_time_Profiles}
    
    The 3D modelling output of EUHFORIA allows us to extract the 1D time series at any time and at any position in the modelling domain, providing us with the possibility to directly compare modelling results with in situ observations. We consider the 1D radial speed and density time series for virtual spacecraft positioned at radial distances between 0.2 and 1.1~au, with a step of 0.1~au. Virtual spacecraft are positioned in the grids with 4$^{\circ}$ spacing (see Fig.~\ref{Fig:motivation_plots}c). We focus our analysis on the subset of virtual spacecraft. We consider virtual spacecraft spanning $\pm$16$^{\circ}$ in latitude on two meridional planes: a) the one containing the Sun-Earth line \citep[as in][]{2024Valentino}, and b) the one containing the main direction of propagation (MDP).
    (The MDP is defined as the line extending radially from the inner to the outer boundary of the domain, along the latitude and longitude prescribed by the CME model, see Fig.~\ref{Fig:motivation_plots}d). 
    Figures~\ref{fig:December_vs_maindir}~to~\ref{fig:October_vs_sunearth} show 1D time profiles for the two studied events. In every panel, the time series along the virtual spacecraft located along the MDP or the Sun-Earth line is shown with a black dashed curve, while the virtual spacecraft above and below these lines appear in shades of blue and red as it is shown in Fig.~\ref{Fig:motivation_plots}d). To capture as many details as possible for each radial distance, we adapted the range of the y-axis in each panel so that the amplitude of the shock covers as much area of the panel as possible. This means that the panels have different ranges on their y-axis, which we indicate with the coloured bars on the right of each panel (for the range of values, see figure captions).
    
    We note that in the case of non-magnetic CMEs, such as the cone model used in this study, the modelled CME profile is often approximated as the shock wave profile and compared to the shock signatures in the in situ observations \citep[which was also the case in][]{2024Valentino}. Herein, we consider the 1D time profiles as the CME front and not the shock wave. The modelled structure, which confines the shock wave conditions, will be discussed in Section~\ref{sect:CME_driven_shock_structure}.

        \subsubsection{Time profiles for Event 1: December~7~2020}
        
        \begin{figure*}
            \centering
            \includegraphics[scale=0.24]{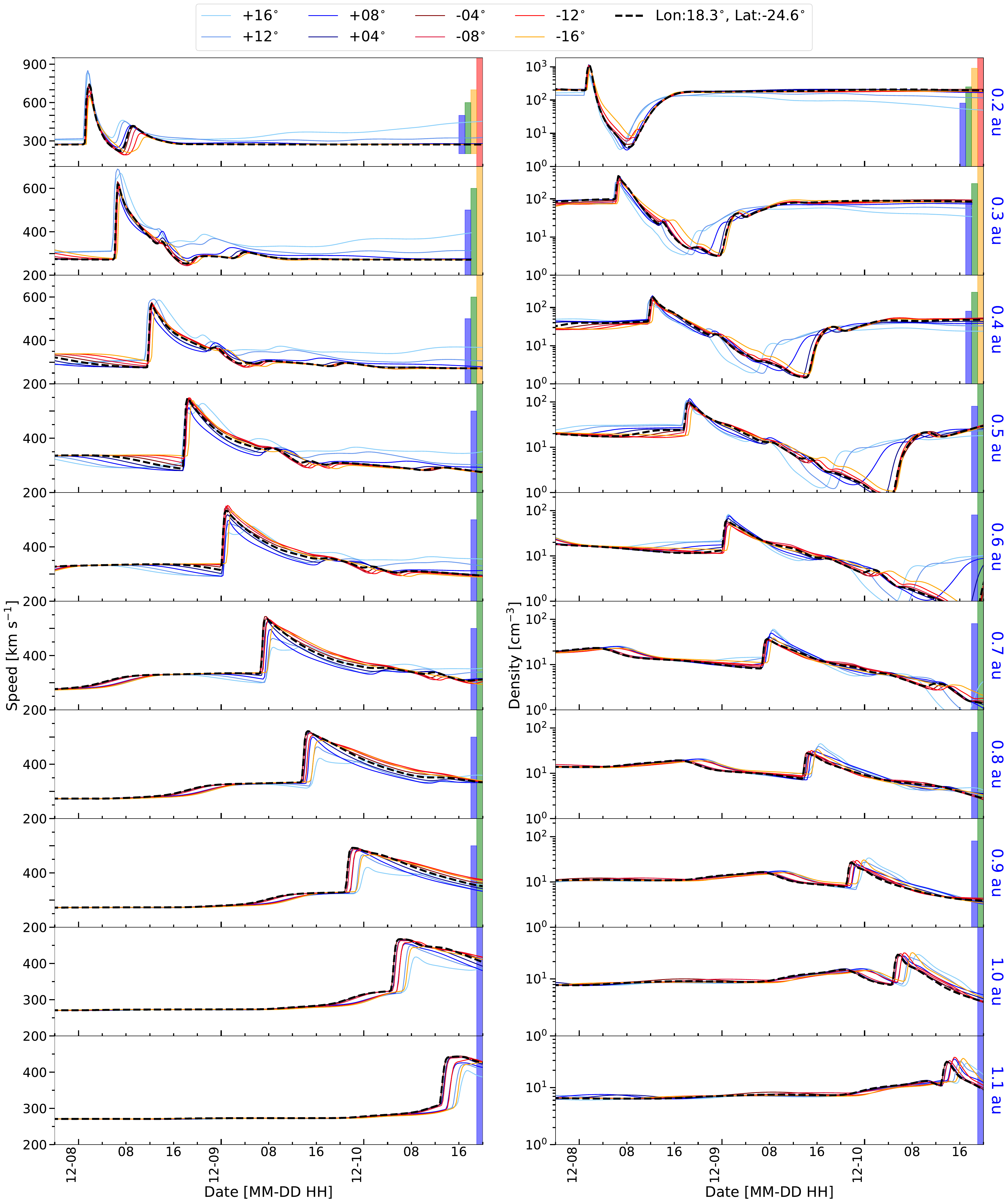}
            \caption{Comparison of the modelled speed (left column) and density time series (right column) for Event~1 \textbf{on the meridional plane containing the MDP} (at +18.3$^{\circ}$ longitude and -24.6$^{\circ}$ latitude).
            Each row of panels shows the results at different radial distances, from 0.2~au up to 1.1~au. MDP is shown in a black dashed line, while the time series at different virtual spacecraft are shown in shades of red (below MDP) and blue (above MDP). The coloured bars on the right side of each panel represent the scale of the y-axis, which is adapted to different ranges to capture the full profile details at different distances. Following ranges are considered: a) Red marks range 100--950~km~s$^{-1}$ for speed and, 1--1850~cm$^{-3}$ for density; b) Yellow marks 200--700~km~s$^{-1}$ for speed and, 1--700~cm$^{-3}$ for density; c) Green refers to range 200--600~km~s$^{-1}$ for speed and, 1--150~cm$^{-3}$ for density; d) Blue refers to 200--500~km~s$^{-1}$ for speed and, 1--80~cm$^{-3}$ for density.}
                \label{fig:December_vs_maindir}
        \end{figure*} 
        
        We first consider the 1D time profiles on the meridional plane containing the \textbf{main direction of propagation (MDP)}, which is about -24$^{\circ}$ in latitude and +18$^{\circ}$ in longitude from the Sun-Earth line (see also Tab.~\ref{tab:Run_Parameters}). Figure~\ref{fig:December_vs_maindir} shows the modelled 1D time series for Event~1 on this plane. The 1D time series shows clear CME signatures in both radial speed and density, already in the first set of virtual spacecraft. The steep speed time profile, which is at 0.2~au rather thin in the radial direction, becomes more stretched and of lower amplitude as we go to the larger distances from the Sun. The CME front, which has a velocity of about 800~km~s$^{-1}$ at 0.2~au, is followed by a dip (down to about 250~km~s$^{-1}$) which lasts about 8 hours and by the second smaller peak of about 400~km~s$^{-1}$. The dip and the second smaller peak are not so clearly identifiable already at 0.3~au, as this trailing part of the CME shows a complex wavy structure. The maximum velocity of the CME front drops to about 650~km~s$^{-1}$ already at 0.3~au. As the distance increases, the speed of the CME front continuously decreases, and the full CME passage lasts longer, from the initial duration of about 8~hours and $\sim$800~km~s$^{-1}$ at 0.2~au to more than 24~hours and $\sim$450~km~s$^{-1}$ at 1.1~au. We note that at 0.2~au the CME time profiles are very similar at all considered latitudes, i.e., different virtual spacecraft. As we go away from the Sun, the CME profiles at different latitudes start to diversify from each other, with a visible decrease in the speed amplitude for the case of virtual spacecraft above the main propagation direction. The peak of the speed amplitude at 1.0~au is at +16$^{\circ}$ lagging about 4~hours behind the peak at MDP and the peaks at latitudes below the MDP.
        
        The 1D density profiles show an initial increase at the CME front when arriving at virtual spacecraft, followed by a density drop, i.e., a trailing low-density region. We note that the cone CME propagation results in the piling up of the ambient plasma ahead of it. The density drop behind the CME appears as time is needed in the 3D modelling domain for the background wind behind the CME to be replenished. 
        This dip becomes more prominent and longer lasting as the CME propagates away from the Sun. Similar to the speed, the density peak is also followed by the increasingly wavy structure of the dip, indicating complex dynamical processes. The virtual spacecraft situated above the MDP, indicated with the blue shaded lines, reach their minimum value within the deep and then return to the background solar wind values faster than the rest of the virtual spacecraft density profiles. This drop in both analysed quantities is associated with the propagation of the CME in the EUHFORIA background solar wind. As the CME propagates faster than the solar wind, it piles up the solar wind plasma immediately in front of itself, leaving a less dense region behind. During its propagation, the CME also expands, which means that the region with the displaced plasma also increases, resulting in the slow restoration of the background solar wind conditions in the larger area.
        
        Before the arrival of the CME front, we can also see the small speed increase (up to about 320~km~s$^{-1}$) consistently across all the virtual spacecraft, visible clearly starting from about 0.7~au. This characteristic appears due to the CME crossing the part of the HCS, which is modelled by EUHFORIA as a disconnected structure. The closed part of the HCS is shown as a pale blue region of a drop-like shape in Fig.~\ref{Fig:Default_EUHFORIA_Dec2020}. The plasma within this region is of a slightly larger density, which is also visible from the density time profiles (Fig.~\ref{fig:December_vs_maindir}). We believe that this increased density is the effect of having plasma in the closed structure (which is most probably a modelling artefact). We note that only from the time profiles, this structure could be misinterpreted as a solar wind flow.

            \begin{figure*}
                \centering
                \includegraphics[scale=0.24]{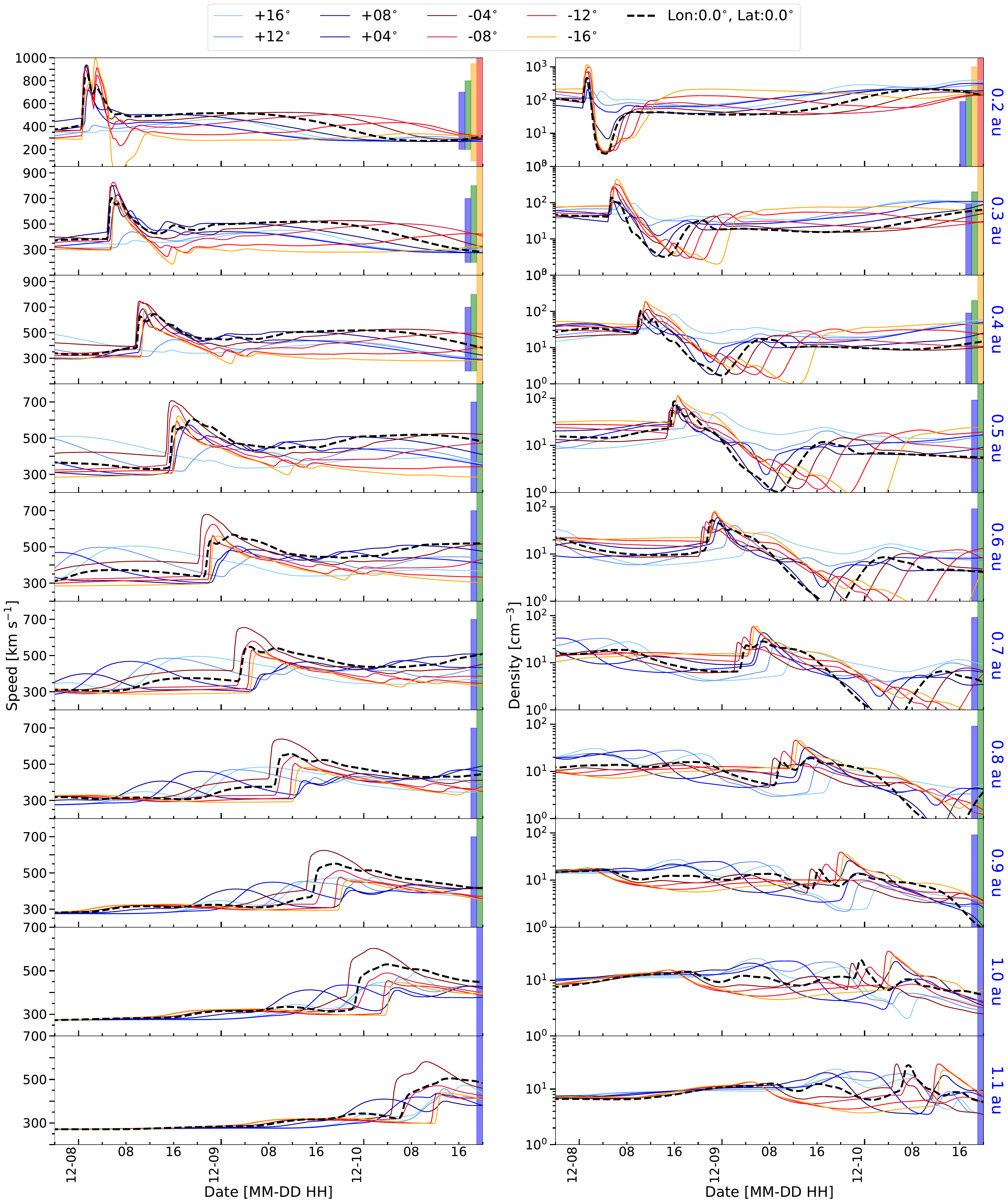}
                \caption{Comparison of the modelled CME speed and density time series (left and right panels, respectively) for Event~1, \textbf{on the meridional plane containing the Sun-Earth line}.
                Each row of panels shows the results at different radial distances, from 0.2~au up to 1.1~au. The Sun-Earth line time series are presented with a black dashed curve. Time series at different virtual spacecraft are shown in shades of red and blue (below and above the Sun-Earth line, respectively). The coloured bars on the right-hand side of each panel represent the scale of the y-axis, which is adapted to different ranges to capture more details at all distances. The coloured bars here correspond to the following ranges: a) Red marks range of 50--1000~km~s$^{-1}$ for speed and 1--1850~cm$^{-3}$ for density; b) Yellow marks 100--950~km~s$^{-1}$ for speed and 1--1000~cm$^{-3}$ for density; c) Green refers to the range of 200--800~km~s$^{-1}$ for speed and 1--200~cm$^{-3}$ for density; d) Blue marks range 200--700~km~s$^{-1}$ for speed and 1--90~cm$^{-3}$ for density.}
            \label{fig:December_vs_sunearth}
            \end{figure*}
        
        After discussing the 1D time profiles on the meridional plane containing the MDP, we focus on those \textbf{on the meridional plane containing the Sun-Earth line}, presented in Fig.~\ref{fig:December_vs_sunearth}. 
        The main difference between the profiles on the two planes is the much larger complexity of the time profiles at different latitudes on this plane.One of the reasons for this complexity is the characteristic that the Sun-Earth line is along the CME flank. The 1D time profiles (for both speed and density) along the Sun-Earth line show that the CME does not arrive simultaneously at all latitudes. 
        This discrepancy in the arrival time of the CME was almost not present when the time series on the meridional plane containing the MDP were considered (Fig.~\ref{fig:December_vs_maindir}). For the MDP, this effect became noticeable rather far from the Sun, at around 0.8~au. The time series at the virtual spacecraft at +16$^{\circ}$ in latitude above the Sun-Earth line do not show a clear signature of the CME front (Fig.~\ref{fig:December_vs_sunearth}). This is in accordance with Figs.~\ref{Fig:Default_EUHFORIA_Oct2021} and \ref{Fig:2021Event_with_arrows}, which show that the CME propagates mainly below the Sun-Earth line. The time profiles at +12$^{\circ}$ in latitude above the Sun-Earth line show only a weak signature of the CME arrival from 0.3~au onwards. These two are the virtual spacecraft furthest from the main direction of propagation, and consequently, the CME signatures are the weakest pronounced at these latitudes. The solar wind speed time profiles rise from the south (red curves) already at 0.2~au, visible as the second peak in the time profiles. This increase in the wind speed is also visible in the meridional cut of Fig.~\ref{Fig:Default_EUHFORIA_Oct2021} as the elevated solar wind speed in front of the CME. The spread in the modelled CME arrival time is larger when is observer further away from the CME nose. This result emphasizes how variations in the direction of propagation of modelled CMEs can vastly affect space weather forecast accuracy.
        
        The background solar wind shows the existence of the solar wind stream SWS~1 in front of the CME, starting to be visible in Fig.~\ref{fig:December_vs_sunearth} from 0.5~au, observed first at the latitudes above Earth, and beyond 0.5~au also at the Earth's latitude. At about the same distance, we also start to notice the change of the CME front profile below the Sun-Earth line (red-shaded curves), which becomes more and more structured, and the spread of different latitude profiles below the Sun-Earth line over time is increasing. 
        
        The CME front shows two nearby maxima in the speed profile at a distance of 0.2~au, indicating a structured CME surface. The dip following the maxima of velocity is more noticeable closer to the Sun and at negative latitudes. As the CME propagates through the modelling domain, this velocity dip gets smoother and less pronounced. Moreover, due to consistently faster ambient solar wind in front of the CME front, the virtual spacecraft positioned at -4$^{\circ}$ (below the Sun-Earth line) shows the earliest arrival time. This is a clear effect of the CME deformation induced by its interaction with the variable solar wind. This effect is also visible in Fig.~\ref{Fig:2020Event_with_arrows} where the -4$^{\circ}$ latitude region is marked by the purple arrow. The speed time profiles also show a clear presence of the fast wind SWS~2 in front of the CME, starting from the latitudes above Earth at distances of already 0.4~au and being more pronounced as we go away from the Sun. The CME fully enters this wind stream approximately at 1~au.

        We found that the 1D density time series show complex profiles (Fig.~\ref{fig:December_vs_sunearth}), similarly to the speed ones. The CME front is followed by the low-density region described earlier (Fig.~\ref{fig:December_vs_maindir}), which stretches in time and distance as we go away from the Sun. The largest differences in the density time profiles between the virtual spacecraft at different latitudes are at 0.1~au.
        
        
        Comparison of the time profiles in the meridional plane containing the Sun-Earth line, with profiles on the plane containing the MDP, shows that after the CME front passage, the values for both speed and density at the different virtual spacecraft do not converge to similar values, but they actually differ quite substantially. In particular, after distances of about 0.6~au, the time profiles become very complex and do not allow us to deduce the structure of the CME, clearly indicating the need for 2D and 3D visualisation. Such a behaviour allows us to conclude that, in this event, 1D time series show how different the speed and density time profiles can be at rather close positions in space. Such an approach is necessary to compare modelling results with observed 1D in situ time series.

        \subsubsection{Time profiles for Event 2: October~28~2021}
    
        \begin{figure*}
            \centering
            \includegraphics[scale=0.24]{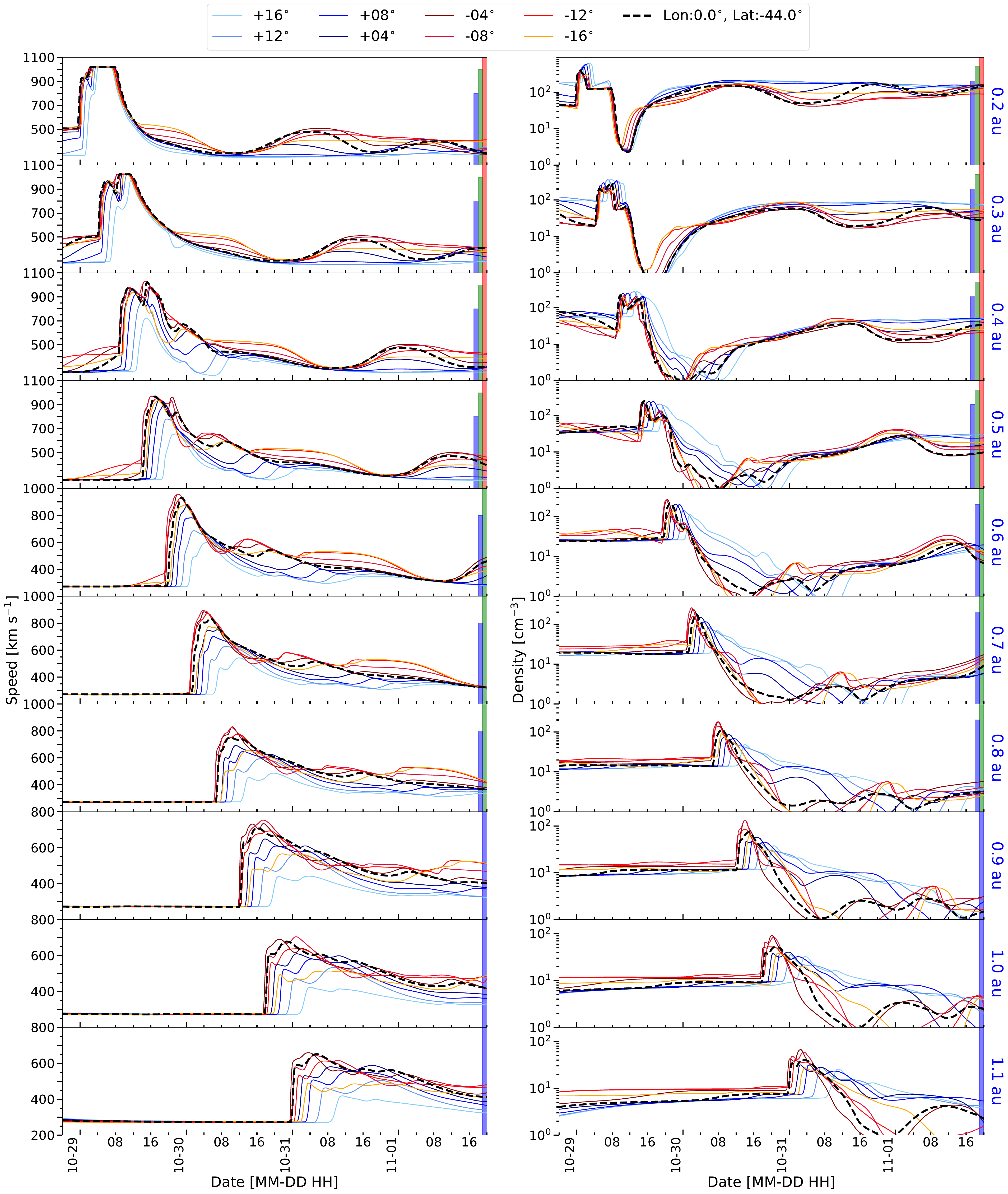}
            \caption{Comparison of the modelled speed (left column panels) and density time series (right column panels) for Event~2 \textbf{in the meridional plane containing the MDP} at 0$^{\circ}$ longitude and -44.0$^{\circ}$ latitude. 
            Each row of panels shows the results at different radial distances, from 0.2~au up to 1.1~au. MDP is shown in a black dashed line, while the different virtual spacecraft are shown in shades of red (below MDP) and blue (above MDP). The coloured bars on the right side of each panel represent the scale of the y-axis, which is adapted to different ranges to capture more details of all the distances. The coloured bars to the right of the figure correspond to the following ranges: a) Red is from 200--1100~km~s$^{-1}$ for speed and from 1--900~cm$^{-3}$ for density; b) Green is from 200--1000~km~s$^{-1}$ for speed and from 1--500~cm$^{-3}$ for density; d) Blue is from 200--800~km~s$^{-1}$ for speed and from 1--200~cm$^{-3}$ for density.}
            \label{fig:October_vs_maindir}
        \end{figure*}
    
        The 1D time profiles for Event 2 \textbf{on the meridional plane containing the MDP} are presented in Fig.~\ref{fig:October_vs_maindir}. In this case, the MDP is located at -44${^\circ}$ latitude and 0${^\circ}$ longitude. As in the previous event, the CME front has clear signatures in both speed and density. The speed time profiles at the closest distance to the Sun of 0.2~au show that the CME front reaches almost simultaneously all virtual spacecraft. As we go to larger distances, the virtual spacecraft that are above the MDP, blue shades, start to diversify, with the profiles at +16$^{\circ}$ changing fastest. The time profiles at the central virtual spacecraft remain up to 0.6~au very similar to the one at latitudes below MDP. Only the profiles at -16$^{\circ}$ (yellow line) start to show late arrival of the CME front beyond 0.7~au. The time series at the virtual spacecraft at +16$^{\circ}$ is the one that consistently arrives the latest at all distances, and it shows the smallest speed profile amplitude. This is in contrast with Event~1, for which the CME was arriving almost simultaneously at all virtual spacecraft and started to show delays in the time profiles only after 0.8~au. This is due the state of the ambient plasma in the simulation domain, which is much more variable and dynamic than in a case of Event~1.

        In this event, the initial peak of the solar wind speed time profile indicates that the CME front has a value of about 950~km~s$^{-1}$. The speed peak is followed by a small dip, then a rise, and then a plateau, a numerical effect of the CME insertion. This plateau has the speed of the inserted CME (1005~km~s$^{-1}$, see Tab.~\ref{tab:Run_Parameters}) and it is observed simultaneously for all virtual spacecraft. The plateau gradually changes to the second peak, which then slowly becomes less pronounced and disappears after 0.6~au. This characteristic is associated with the size of the cone CME injected into EUHFORIA's heliospheric domain, which in the case of this event is relatively large and amounts to about 98$^{\circ}$. For the case of wide CMEs, the amount of time the model is injecting the cone CME exceeds the time it takes the nose of the CME to reach 0.2~au, and therefore, the CME parameters are still fixed by the CME's speed and density, and not by the MHD model. In the case of the density time profiles, the plateau is not as well defined as in the case of the CME speed, but it is still present. Similarly to the plateau in the speed, it's formed at lower values (200~cm$^{-3}$) than the density peak, which indicates CME arrival. It also transforms to the second peak, which disappears after 0.6~au. Similarly to Event~1, the density drop which follows the CME passage deepens and becomes wider in a case of larger distances from the Sun.
    
         We note that up to distances of 0.6~au, a stream of fast solar wind is observed in front of the CME, which is then fully encompassed by the CME. After the passage of the CME front, for both speed and density, time series at different virtual spacecraft reach different values, indicating passage through the non-uniform solar wind. The time series below the MDP (shades of red) map higher speeds as they are passing through a structured stream of fast solar wind.

        \begin{figure*}
            \centering
            \includegraphics[scale=0.24]{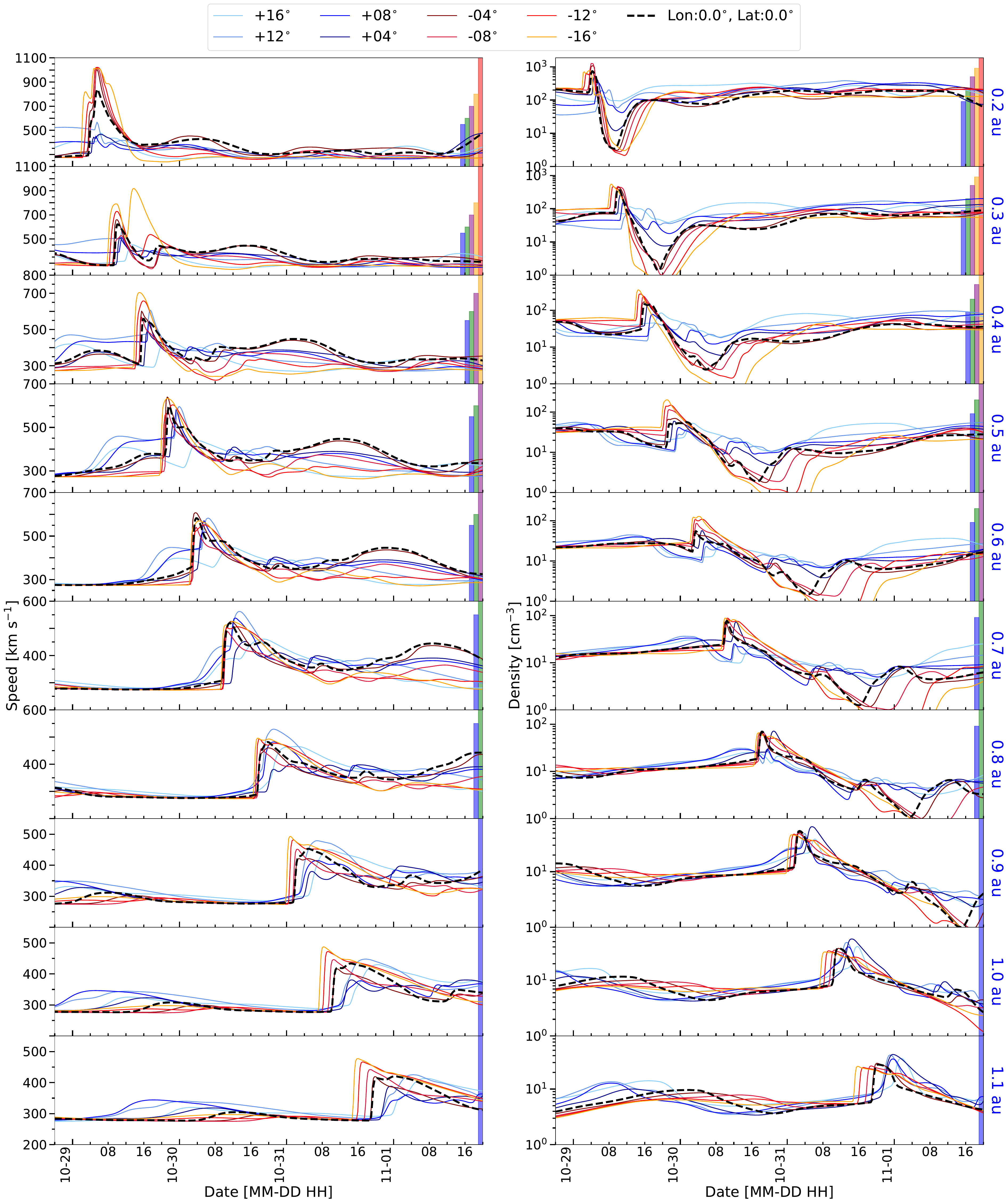}
            \caption{Comparison of the modelled speed (left column panels) and density time series (right column panels) for Event~2 \textbf{in the meridional plane containing the Sun-Earth line}. 
            Each row of panels shows the results at different radial distances, from 0.2~au up to 1.1~au. The Sun-Earth line is shown in a black dashed line, while the different virtual spacecraft are shown in shades of red (below the Sun-Earth line) and blue (above the Sun-Earth line). The coloured bars on the right side of each panel represent the scale of the y-axis, which is adapted to different ranges to capture more details of all the distances. The coloured bars here correspond to the following ranges: a) Red is from 200--1100~km~s$^{-1}$ for speed and from 1--1850~cm$^{-3}$ for density; b) Orange is from 200--800~km~s$^{-1}$ for speed and from 1--900~cm$^{-3}$ for density; c) Purple is from 200--700~km~s$^{-1}$ for speed and from 1--500~cm$^{-3}$ for density; d) Green is from 200--600~km~s$^{-1}$ for speed and from 1--200~cm$^{-3}$ for density; e) Blue is from 200--550~km~s$^{-1}$ for speed and from 1--90~cm$^{-3}$ for density. } 
                \label{fig:October_vs_sunearth}
        \end{figure*}
    
        Figure~\ref{fig:October_vs_sunearth} shows the simulated in situ time series for Event~2, \textbf{in the meridional plane containing the Sun-Earth line}. The Sun-Earth line is 44$^{\circ}$ above the MDP of the CME, which results in, similarly to Event~1, a transition through the CME flank. Since the virtual spacecraft are positioned at maximum $\pm$16$^{\circ}$, there is no overlap between them. As in the case of Event~1, we also see the complexity of the solar wind modelled in the time series at spacecraft positioned at different latitudes. Although the CME front arrives at the same time at all virtual spacecraft, the speed profiles at 0.2~au show different maximum values. Those above Earth (blue shades) show a very small speed increase, while the ones below Earth have values at almost the insertion speed (1005~km~s$^{-1}$). This behaviour follows from the characteristic that MDP is also at latitudes below the Sun-Earth line. At larger distances, due to the interaction with the ambient solar wind, the differences between the CME front amplitudes are smoothed out. 
        Similarly to the modelling results of the time series on the meridional plane containing the MDP (Fig.~\ref{fig:October_vs_maindir}), we notice the presence of a fast solar wind stream before and around the arrival of the CME front. A second peak in speed, due to interaction with the SWS1, appears at 0.2 and 0.3~au for the time series at virtual spacecraft below Earth. The time series that shows the strongest second peak in speed is the one for the virtual spacecraft positioned at -16$^{\circ}$, with values close to $\sim$900~km~s$^{-1}$, which disappears after 0.4~au. The CME gradually catches up to this wind flow, and they fully merge close to 0.7~au. Another solar wind flow, SWS2, observed in front of the CME front (see also Figs.~\ref{Fig:Default_EUHFORIA_Oct2021} and \ref{Fig:2021Event_with_arrows}), starts to be visible at about 0.9~au, also first in the time series of the virtual spacecraft above Earth. We would like to note that after 0.9~au, due to interaction with the variable solar wind, the differentiation in arrival time of the CME front between the virtual spacecraft becomes significantly large (at 1.1~au it is about 8 hours).
    
        In the case of the density time profiles, the virtual spacecraft above Earth shows very low amplitudes, i.e., not well pronounced signatures for the CME front arrival up to distances of about 0.4~au. The time series at the virtual spacecraft below the Sun-Earth line shows very similar maximum values for the full distance of 0.2-1.1~au. The density profiles show a dip after the arrival of the CME front, which gets stretched out as the CME propagates to larger distances. Similar behaviour was observed for all density time profiles and directions of propagation (see Figs.~\ref{fig:December_vs_sunearth}, \ref{fig:October_vs_maindir} and \ref{fig:October_vs_sunearth}).

\section{Structuring of the CME-driven shock wave}
\label{sect:CME_driven_shock_structure}

Propagation of the CME through the solar corona can cause the large amplitude perturbation in the ambient plasma, which then can transform into a shock wave. However, the necessary conditions for the formation of the CME-driven shock wave are not always met. Moreover, the large amplitude waves might not fulfil the shock conditions (Sect~\ref{sect:Shock_tracing_method}) simultaneously in front of the whole CME \citep[see e.g., ][and references therein]{Mann95, Vrsnak08, Bacchini15}. 

Additionally, during its propagation, the CME can undergo different transformations, e.g., it expands and decelerates. Such CME changes can also reflect on the CME-driven shock wave. It is generally considered that the CME nose (i.e., the region closer to the MDP) propagates the fastest and drives the strongest part of the shock. The probability to fulfil the shock wave conditions then gradually decreases towards the CME flanks. This characteristic has implications for different phenomena linked to interplanetary shocks, e.g., acceleration and propagation of SEPs \citep{2023Wijsen}. On the other hand, it is necessary to note that the shock-associated radio emission, i.e., type II radio bursts are often observed at close to the CME flank regions. That association is considered to be related to the frequently present magnetic field geometry favourable for the generation of radio emission \citep{Magdalenic02}, close to the CME flank regions, and in the case of the existence of nearby streamers \citep[see e.g., ][]{Magdalenic14, Zucca18}. 

While previously we discussed propagation of the CME front modelled with the cone CME model (Sect.~\ref{sect:modelling_3D_2D_1D}), in this section we address the associated structure which fulfils the Rankine-Hugoniot relations, i.e., CME-driven shock wave. 
Both of the herein studied events were associated with the type II radio bursts \citep{2024Valentino}, which observationally confirms the presence of the CME-driven shock wave. As the Rankine-Hugoniot conditions depend on both upstream and downstream solar wind plasma characteristics, the variable background solar wind is expected to influence deformations of the CME-driven shock wave.

\begin{figure*}[h!]
        \includegraphics[width=0.98\textwidth]{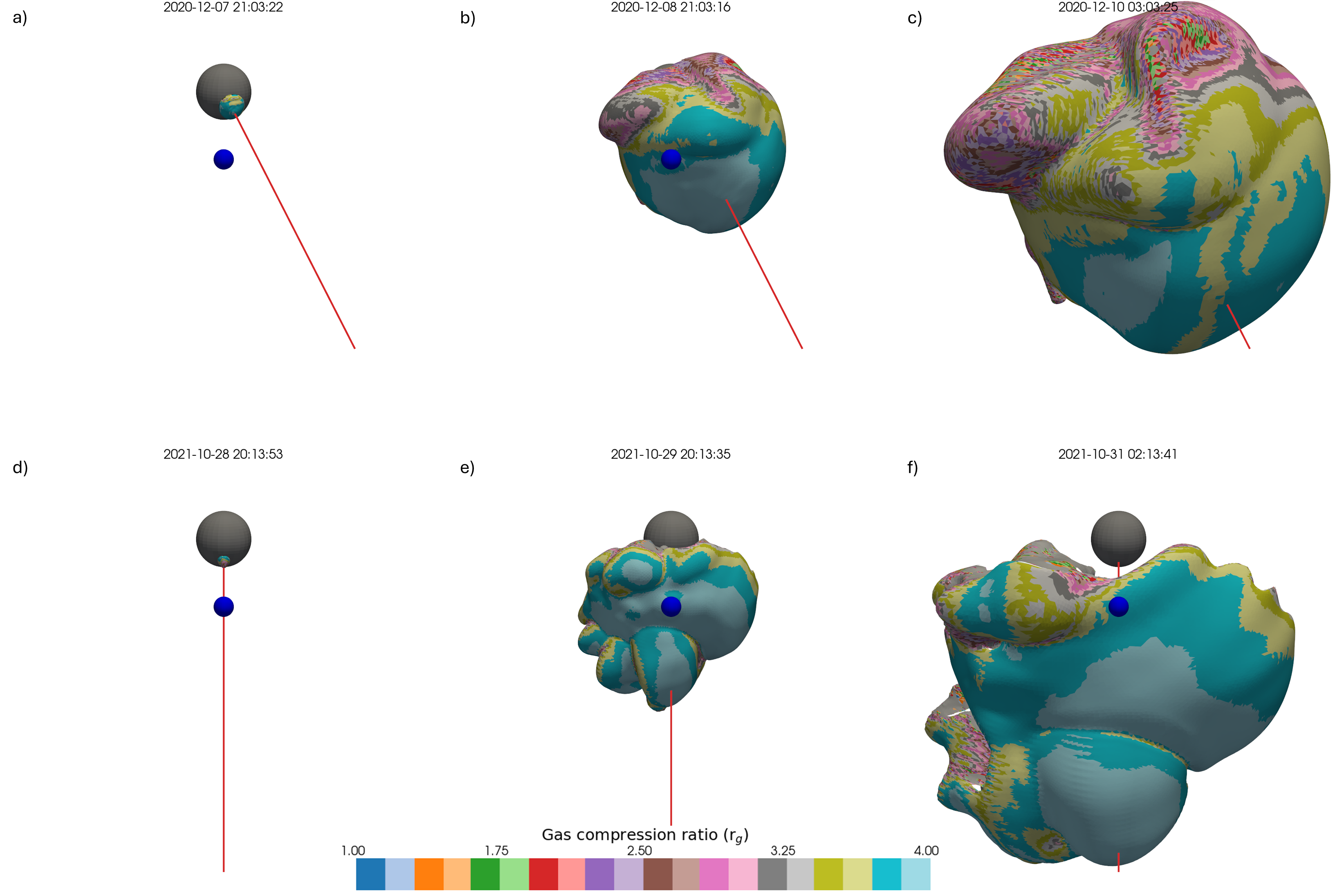}
        \caption{CME-driven shock surface as extracted by the shock tracing algorithm at three different times for Event~1 (panels~a~to~c) and Event~2 (panels~d~to~f). The MDP of the CME is represented with a red line, the inner boundary is represented with the dark grey sphere, and Earth position is mapped with the ink-blue sphere. The colour scale of the shock front maps the shock compression ratio ($r_{g}$). Panels a and d correspond to the first instant in which the shock is clearly visible in the 3D presentation. Panels b and e correspond to a time where the shock is at a distance of $\sim$0.5~au, in the direction along the MDP. Panels c and f correspond to a time where the CME is $\sim$1~au in the direction along the MDP.}
        \label{Fig:Gas_Comp_Ratio}
    \end{figure*}

As a first step, we locate the exact regions in which the CME is driving a shock wave using the method described in Sect.~\ref{sect:Shock_tracing_method}. The mapped shape is a strongly structured CME-driven shock surface, which has a 'knobbly ginger-like' shape. Structured shock is shown in the panels~a~to~c and panels~d~to~f of Fig.~\ref{Fig:Gas_Comp_Ratio} for Event~1 and Event~2, respectively. We show the evolution of the CME-driven shocks at three different times. A grey sphere, at the centre of the domain, represents the inner boundary of EUHFORIA, while Earth is represented by a small ink-blue sphere (Earth's size is not up to scale). The main direction of propagation of the CME is marked with a red line. The panels~a and d show the CME-driven shock at the first instance (for Event~1 and 2, respectively), when its structure becomes fully and clearly visible in the 3D presentation. Panels b, e, and c, f correspond to snapshots in which the shock is along the MDP at distances of approximately 0.5 and 1~au, respectively. 
    
The shock wave front shows a smoother structure and a more spherical shape in the case of Event~1 (panels~a, b, and c) than in the case of Event~2 (panels~d, e, and f). Different structuring of the shock front is related to the different background solar wind in which the events are propagating. In the case of Event~1, two solar wind streams SWS~1 and SWS~4 seen close to the northern flank of the CME, and the overall not strongly structured background solar wind conditions influence the change of the shape of the expanding CME-driven shock (see also Fig.~\ref{Fig:2020Event_with_arrows}). Panel~b shows the formation of a knob, i.e., a circular, rounded bulge, on the northern section of the CME-driven shock wave surface. The position of this knob roughly coincides with the position of SWS1, indicating the faster shock wave expansion at these latitudes due to interaction with the faster solar wind. We note that, even if the protruding part of the shock front does not change much between the two presented time instances (panels b and c of Fig.~\ref{Fig:Gas_Comp_Ratio}), the gas compression ratio, which relates to the shock wave strength, decreases from about 3.85 to about 3.00. At lower latitudes and closer to the main direction of propagation, the shock surface remains quite uniform and smooth, and it expands relatively symmetrically, but as the CME expands and decelerates, the CME-driven shock wave weakens, i.e., the gas compression ratio decreases from about 4.00 to 3.55. 
The northernmost part of the CME-driven shock weakens the fastest, with the gas compression ratios dropping from close to 3.25 when the shock is first mapped to close to 2.50 when the shock is close to 1~au. Along the MDP, the CME-driven shock also weakens, but significantly less than along the Sun-Earth line, with the gas compression ratio of approximately 4.00 decreasing to below 3.50.

In the case of Event~2, the structure of the CME-driven shock front is significantly more complex than in the case of Event~1. This results from the strongly complex and convoluted general structure of the background solar wind modelled by EUHFORIA (see Figs.~\ref{Fig:Default_EUHFORIA_Oct2021} and \ref{Fig:2021Event_with_arrows}). The MDP of the CME is along a region with several narrow solar wind streams, which highly influences the shape of the CME-driven shock of Event~2. This part of the shock eventually evolves into a few knobs with two of them being more prominent (panels~e and f).  The shock weakening seen through the decrease of the gas compression ratio is not as strong as in the case of Event~1. In the northern flank regions, the gas compression ratio decreases from about 4.00 to values close to 3.00. Along the MDP, rather small changes in the shock wave strength are observed. For comparison, in the case of Event~2, the CME-driven shock surface starts to visibly fragment as a decreasingly smaller part of the eastern flank region satisfies the Rankine-Hugoniot conditions. Consequently, the automatic tracing algorithm stops detecting this region as part of the CME-driven shock. In the case of Event~1, such regions are small and barely observable in the flank regions.

\section{Evolution of the CME-driven shock wave}
\label{sect:evolution_CME_Structure}
    Using the shock tracing algorithm described in Sect.~\ref{sect:Shock_tracing_method}, we followed the evolution of the CME-driven shock throughout the simulation domain starting at the time of the CME injection in the EUHFORIA's heliospheric domain. Figures~\ref{Fig:Dec20_Delta_distance} and~\ref{Fig:Oct21_Delta_distance} show two sets of time series (panels a and b) and a visualisation of the position of the CME-driven shock at three different times for Event~1 and Event~2, respectively. The Earth is marked as a blue sphere, and EUHFORIA's inner boundary as a grey sphere situated in the centre of the domain.
    Panel~a top shows $\Delta\mathrm{Distance}_{i}$ for the Sun-Earth direction, defined as:
    \begin{equation}
        \Delta\mathrm{Distance}_{i} \equiv X_{\mathrm{Sun-Earth}} - X_{\mathrm{pos_{i}}}
        \label{eq:delta_distance}
    \end{equation}
    and in panel~a bottom $\Delta\mathrm{Distance}_{j}$ for the MDP direction, defined as:
    \begin{equation}
        \Delta \mathrm{Distance}_{j} \equiv X_{\mathrm{MDP}} - X_{\mathrm{pos_{j}}}
        \label{eq:delta_distance_MDP}
    \end{equation}
    where $X_{\mathrm{Sun\text{-}Earth}}$ denotes the heliocentric distance of the shock surface along the Sun–Earth line, and $X_{\mathrm{pos}_{i}}$ corresponds to the heliocentric distances of the shock surface along a spectrum of directions at different angular separations ($i$). The bottom sub-panel of panel~a shows the same $\Delta\mathrm{Distance}$, but calculated with respect to the MDP and not the Sun–Earth line (Eq.~\ref{eq:delta_distance_MDP}), different angular separations are here defined as ($j$). Panel~b displays the shock wave's radial distance from the Sun, i.e., $X_{\mathrm{pos_{i}}}$ and $X_{\mathrm{pos_{j}}}$ as a function of time since the CME injection along the Sun–Earth and the MDP directions in the top and bottom sub-panel, respectively. For easier understanding of panels a and b, we show in panel~c three snapshots of the shock wave surface as viewed from the side (see also Fig.~\ref{Fig:Gas_Comp_Ratio}). The lines corresponding to the different angular separations are also plotted. Solid and dashed lines indicate angular differences with respect to the Sun–Earth and the MDP line, respectively.

    \subsection{Shock wave in Event 1}
        \begin{figure*}[h!]
            \includegraphics[width=0.98\textwidth]{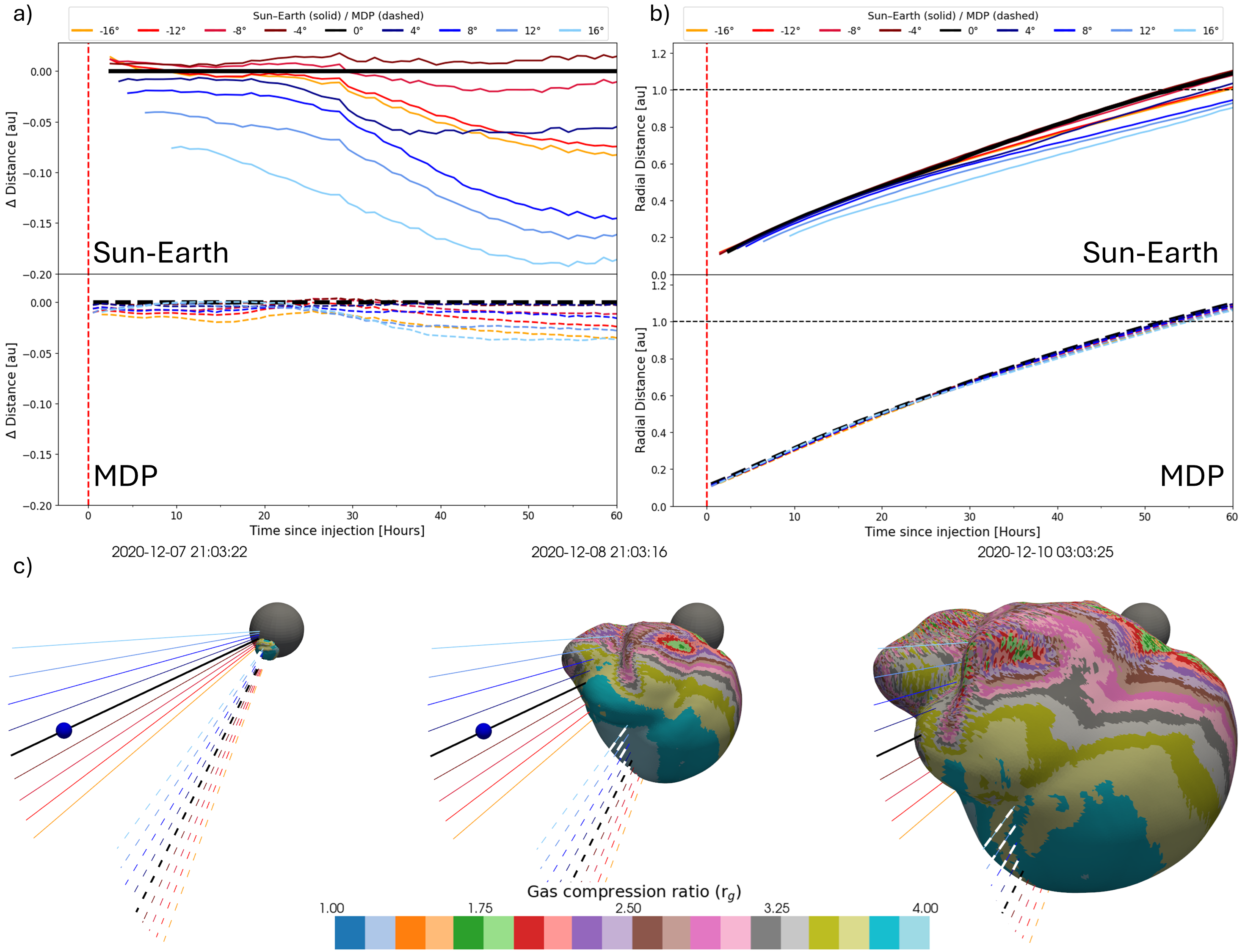}
            \caption{Time evolution of $\Delta$Distance with respect to the Sun-Earth line (top of the panel~a) and with respect to the MDP (bottom of the panel~a) along the different angular separation lines. Time evolution of the radial distance of the shock wave from the Sun, along the different angular separation lines for the Sun-Earth ($X_{\mathrm{pos}_{i}}$) and MDP line ($X_{\mathrm{pos}_{j}}$), shown in top and bottom of panel~b (respectively). Shock wave surfaces at three different time instances in the simulation (panel~c), with the different angular separation lines (full and dashed lines in the vicinity of the Sun-Earth and MDP lines, respectively.}
            \label{Fig:Dec20_Delta_distance}
        \end{figure*}
        
Figure~\ref{Fig:Dec20_Delta_distance}a shows a very distinctive difference in the behaviour of the CME-driven shock along the Sun–Earth line (solid lines, top panel~a) compared to the MDP (dashed lines, bottom panel~a). Near the Sun–Earth line, even slight deviations, particularly towards positive latitudes (blue shaded lines), lead to the large differences in the shock position. This difference rapidly increases as the shock propagates away from the Sun, with the effect being particularly strong and visible after 20 hours since the CME injection. 
In the latitudes below the Sun-Earth line, this effect is less prominent and $\Delta$Distance shows the strongest deviation at latitudes -12$^{\circ}$ and -16$^{\circ}$. On the contrary, along the MDP (dashed lines, panel~a), the evolution of the shock remains fairly uniform and consistent throughout the simulation, particularly for the latitudes below the MDP line (shades of red). At this location, the background solar wind is more homogeneous and therefore has less influence on the shock evolution compared to the Sun–Earth line (see also Sect.~\ref{sect:EUHFORIA_default_Dec2020} and Fig.~\ref{Fig:2020Event_with_arrows}). 
At around 25 hours into the simulation (panel~a, bottom), all the latitudinal profiles begin to display a change, which is possibly due to the interaction with the HCS. The higher latitudes start to deviate from the MDP, but since that deviation is less than 0.05~au (panel~a), this deviation does not result in a significant delay in shock arrival at 1~au (panel~b). Such a behaviour is explainable by the nearly spherical expansion of this part of the shock, clearly seen in the 3D presentation (panel~c). We note that, along the Sun–Earth line, the shock is formed progressively later as we move to higher latitudes (blue shaded lines). The line at +16$^{\circ}$ shows the shock forming the latest, nearly 10 hours into the simulation, and exhibits the largest $\Delta$Distance relative to the Sun–Earth line, with differences approaching 0.2~au after $\sim$50 hours. This offset is also shown in the top panel of Fig~\ref{Fig:Dec20_Delta_distance}~b in which the shock's radial distance from the Sun $X_{\mathrm{pos}_{i}}$ is presented as a function of time since the CME injection. 
We also see that such a behaviour of the time series at different latitudes consequently results in different arrival times to 1~au. In particular, the lines above the Sun-Earth one (blue shaded ones), do not even reach the Earth in the considered time domain. We note that similar trend, of the strongest latitudinal variations being more prominent along the Sun–Earth line rather than the MDP, is evident in all panels of Fig.~\ref{Fig:Dec20_Delta_distance}.

As already mentioned, the 3D presentation in panel~c provides additional explanation for the observed behaviour of the time series. The shock front close to the MDP displays a large knob, which is more prominent towards the southern latitudes. This agrees with the almost simultaneous arrival time of all time series (panel~b bottom). On the other hand, the two smaller knobs stretching at different latitudes along the Sun–Earth line show noticeable variability in the shock front arrival time along the Sun-Earth line. These results allow us to better understand the accuracy-variability of forecasting the shock wave arrival at Earth \citep[e.g., ][]{2007Zhang, 2022Maharana, 2024Rodriguez}. Our findings accentuate the importance of the MDP of the CME and CME-driven shock wave, as discussed in \cite{2024Valentino} and their interaction with the background solar wind.

    \subsection{Shock wave in Event 2}
        \begin{figure*}[h!]
            \includegraphics[width=0.98\textwidth]{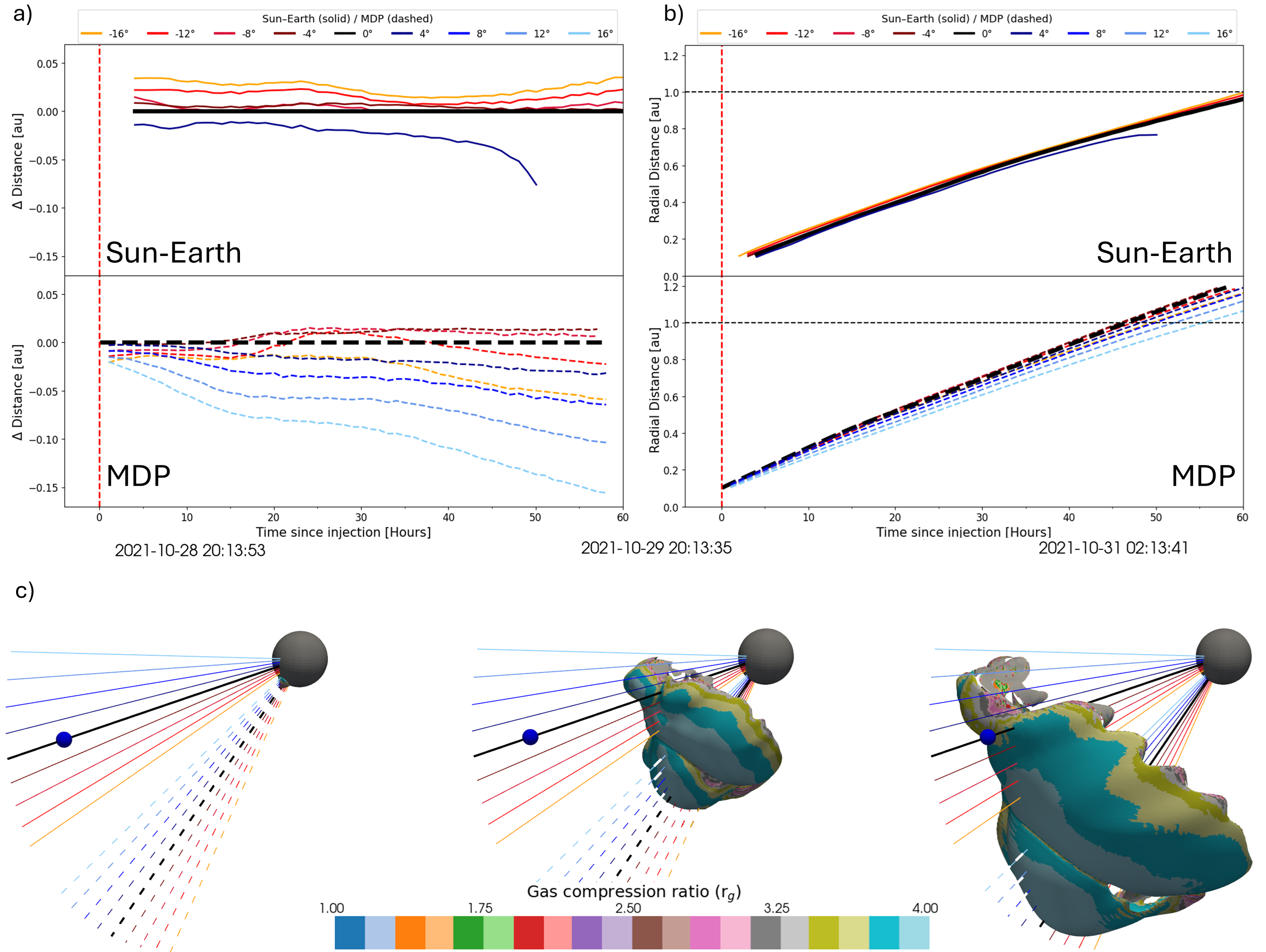}
            \caption{Time evolution of $\Delta$Distance with respect to the Sun-Earth line (top of the panel~a) and with respect to the MDP (bottom of the panel~a) along the different angular separation lines. Time evolution of the radial distance of the shock wave from the Sun, along the different angular separation lines for the Sun-Earth ($X_{\mathrm{pos}_{i}}$) and MDP line ($X_{\mathrm{pos}_{j}}$), shown in top and bottom of panel~b (respectively). Shock wave surfaces at three different time instances in the simulation (panel~c), with the different angular separation lines (full and dashed lines in the vicinity of the Sun-Earth and MDP lines, respectively.}
            \label{Fig:Oct21_Delta_distance}
        \end{figure*}

    The CME injection longitude in the heliospheric part of EUHFORIA (see Tab.~\ref{tab:Run_Parameters}) was for Event~2 0$^{\circ}$. For comparison, the injection longitude for Event~1 was 18.3$^{\circ}$. The main differences among the angular separation lines in the case of Event~2 can be associated with the latitudinal variation of the background solar wind. Along the Sun-Earth line (top panel of Fig.~\ref{Fig:Oct21_Delta_distance} a), the CME-driven shock does not seem to exist above +4$^{\circ}$, and therefore it is not possible to compute the $\Delta$Distance (as defined in Eq.~\ref{eq:delta_distance}). This can be also clearly seen from panel~c, as the shock surface in the northern flank region never intersects the above Sun-Earth lines. For all below the Sun-Earth lines, a delay of $\sim$3~hours in the arrival of the shock structure to these latitudes is observed. The overall behaviour of the below the Sun-Earth lines is similar to $\Delta$~Distance~<~$|0.05|$~au throughout the simulation. The exception is the +4$^{\circ}$ line, which shows a rapid increase after 40~hours since the CME injection, and at $\sim$50~hours the shock stops being mapped.
    
    Along the MDP (bottom of panel~a), the behaviour is quite different from the case of along the Sun-Earth direction. On the other hand, similar behaviour was observed in the case of Event~1 along the Sun-Earth lines. For the positive latitudes, $\Delta$Distance systematically increases as the simulation progresses in time. The effect is particularly strong for +16$^{\circ}$, and somewhat less pronounced as the lines get closer to the MDP. In the case of lines between +4$^{\circ}$ and -12$^{\circ}$ $\Delta$Distance remains within $|0.05|$~au. In particular, along the MDP, the CME-driven shock structure exhibits a prominent knob, as at these latitudes we have the presence of the solar wind flow (see pink arrow in Fig.~\ref{Fig:2021Event_with_arrows}), with +16$^{\circ}$ and -16$^{\circ}$ lines close to the edges of this sub-structure within the shock surface. 
        
    The evolution of the radial distances at which the shock wave appears, along the different latitudinal lines at the Sun-Earth direction (top panel~b), is very similar. The exception is line along +4$^{\circ}$ which exhibits the biggest difference and an abrupt ending at $\sim$50~hours. We attribute this behaviour to the interaction of the CME-driven shock with the HCS. As can be seen in the meridional, equatorial, and transversal cuts of Fig.~\ref{Fig:Default_EUHFORIA_Oct2021} and in the 3D presentation of Fig.~\ref{Fig:2021Event_with_arrows}d, at these latitudes the shock reaches the HCS. The strong background gradients in density across this region disturb the relative-entropy signatures on which the shock-tracing algorithm relies (Sect.~\ref{sect:Shock_tracing_method}), so that the Rankine-Hugoniot conditions are no longer identified there and this portion of the shock ceases to be mapped. This is consistent with the fragmentation of the shock surface in the flank regions discussed in Sect.~\ref{sect:CME_driven_shock_structure}. The shock wave arrives at 1~au more or less simultaneously at $\sim$60~hours after the injection of the CME. On the other hand, along the MDP (bottom panel~b), the CME-driven shock arrives at 1~au $\sim45$~hours after the CME is injected into the simulation, with lines within -16$^{\circ}$ and +4$^{\circ}$ arriving within a 4~hour time-window. For lines above +8$^{\circ}$, the difference is larger, with +16$^{\circ}$ arriving at 1~au more than 10~hours after the MDP. We conclude that the strong structuring of the solar wind along the MDP resulted in a significantly larger spread of values of the shock wave arrival time than in the case of a more homogenous solar wind along the Sun-Earth line latitudes.

\section{Summary, discussion and Conclusions}
\label{sect:Summary_and_conclusions}
    
    \subsection{What are the advantages of combining the 3D, 2D and 1D analysis}
    \label{sect:Summary}
    
    The goal of this study was to understand how the interaction with the ambient solar wind affects the structure and propagation of the CME and CME-driven shock wave. We employed EUHFORIA and the cone model to model two CMEs: Event~1 on December 7, and 2020 and Event~2 on October 28, 2021. In this comprehensive analysis of the MHD simulation results, we link the CME and the CME-driven shock structures to the characteristics of the solar wind as seen in 3D, 2D, and 1D domains. Due to CME's particular direction of propagation and a very different ambient background solar wind, the studied events are excellent examples of how interaction with the ambient wind can modify the structure of the CME and CME-driven shock and how challenging it is to understand their dynamics using 2D plots or time series alone. 
    
    In order to follow the evolution of the CME and the CME-driven shock wave from the point of insertion in the EUHFORIA's heliospheric domain at 0.1~au and up to 1~au, we introduced a new methodology employing 3 different dimensions (3D, 2D, and 1D). We also treated separately the CME from the CME-driven shock wave isolated using the Rankine-Hugoniot conditions. We discuss here the most important components of our multi-dimensional study: 
    
    \begin{itemize}
        \item[a)] The 2D meridional and equatorial cuts from the EUHFORIA simulation results (Figs.~\ref{Fig:Default_EUHFORIA_Dec2020}, \ref{Fig:2020Event_with_arrows},~\ref{Fig:Default_EUHFORIA_Oct2021} and \ref{Fig:2021Event_with_arrows}) are usually the first approach modellers use for understanding the ambient solar wind characteristics in which the CMEs propagate. However, such an approach has some drawbacks. Only the 2D plane cuts do not provide sufficient information on the interaction of the CME and background wind, as some of the solar wind flows that interact with the CME might be observed only in one of the 2D planes (see e.g., Figs.~\ref{Fig:Default_EUHFORIA_Dec2020} and \ref{Fig:2020Event_with_arrows}). Therefore we combined the 2D and 3D presentations of the EUHFORIA's radial solar wind and CME speed and density (Figs.~\ref{Fig:Default_EUHFORIA_Dec2020}, ~\ref{Fig:2020Event_with_arrows}, \ref{Fig:Default_EUHFORIA_Oct2021} and~\ref{Fig:2021Event_with_arrows}). The 3D presentation in the form of isosurfaces, which can be rotated in all directions, complements the 2D plots and allows the accurate identification of the extent of the solar wind flows in different planes. It stipulates an understanding of which structures of the ambient solar wind are really impacting the CME and to what extent. Moreover, by using these two different visualisations, we were also able to isolate and understand the specific features seen in 2D and 3D and connect them to those appearing in the in situ time series, i.e., in 1D presentations.
    
        \item[b)] Understanding 1D presentation is a very important part of this study since this is the form in which the in situ observations, the main and only way to fully validate the accuracy of the modelling, are available. Combining the 3D, 2D, and 1D presentations of the modelling results allows us to have a complete understanding of the overall evolution of the heliospheric environment in the vicinity of the CME, as well as the validation of the modelling accuracy. We employed 1D time series for virtual spacecraft positioned at different radial distances (from 0.2~au to 1.1~au) along the two propagation directions, i.e.,  the MDP and along the Sun-Earth line (Figs.~\ref{fig:December_vs_maindir}, \ref{fig:December_vs_sunearth}, \ref{fig:October_vs_maindir} and \ref{fig:October_vs_sunearth}). Along each of the two selected propagation directions we positioned virtual spacecraft in lines at $\pm$16$^{\circ}$ in order to better understand the variations of the time series and development of the studied events at different latitudes. 
            
        \item[c)] The 3D presentation also allowed us to inspect the structure of the HCS - Heliospheric Current Sheet (Figs.~\ref{Fig:2020Event_with_arrows} d and \ref{Fig:2021Event_with_arrows} d for Events~1 and 2, respectively). The HCS, which provides us information on the shape of the global magnetic field of the Sun, can be very complex and structured, as e.g., in the case of the herein studied Event~2. Since the global magnetic field of the Sun influences both the solar wind and the CMEs, knowledge on HCS shape is very important. We note that even in the case of the cone CME model, modelling results show that the crossing of the HCS impacts the shape of the CME and the CME-driven shock wave (see Fig.~\ref{Fig:2021Event_with_arrows}). However, the physically more correct description of the CME and the HCS interaction is expected to be obtained when simulations are made with the magnetic CME model. This topic will be addressed in the follow-up study in which the FRi3D magnetic CME model is employed (Valentino et al, in prep).
        
    \end{itemize}
    
    We also considered the evolution of the CME-driven shock wave detected with the shock tracing method (Sect.~\ref{sect:Shock_tracing_method}). Such an approach allowed us to accurately locate the position and the characteristics of the CME-driven shock, i.e.,: 
    \begin{itemize}
        \item[$\bullet$] map the surface of the shock wave and the variability of the gas compression ratio $r_g$ along the surface and in different instances in time (Fig.~\ref{Fig:Gas_Comp_Ratio} panels a and c for Event~1, and panels d and f for Event~2),
        \item[$\bullet$] study the time evolution of the relative and absolute position of the shock at different latitudes (panels a and b of Figs.~\ref{Fig:Dec20_Delta_distance} and \ref{Fig:Oct21_Delta_distance} for Events~1 and 2, respectively), 
        \item[$\bullet$] visualize the shock wave deformations at different distances in 3D (as seen from the western side in Figs.~\ref{Fig:Dec20_Delta_distance} c and \ref{Fig:Oct21_Delta_distance} c for Events~1 and 2, respectively),  
        \item[$\bullet$] quantify time differences of the CME-driven shock arrival at different positions at 1~au, with respect to the MDP and Sun-Earth line (summarised in Fig.~\ref{Fig:Time_differentce}). 
    \end{itemize}

    \subsection{Discussion and conclusions}
    \label{sect:Discussion and conclusions}
    
    The 1D profiles of solar wind speed and density extracted from the simulations show marked variations depending on the position of the virtual spacecraft (e.g., Figs.~\ref{fig:December_vs_maindir}–\ref{fig:October_vs_sunearth}). Even when the spacecraft are distributed over a relatively wide range of latitudes and longitudes, the simulation results cannot be fully interpreted from this information alone and therefore require complementary diagnostics. Even when the 1D profiles are accompanied by 2D plots (see Figs.~\ref{Fig:Default_EUHFORIA_Dec2020} and \ref{Fig:Default_EUHFORIA_Oct2021}), as shown in Sect.~\ref{sect:modelling_3D_2D_1D}, the full information is not available and combination may be misleading, particularly when CMEs propagate outside the ecliptic plane. As demonstrated in Figs.~\ref{Fig:2020Event_with_arrows} and~\ref{Fig:2021Event_with_arrows}, only employing the 3D visualisation of the simulation results removes the uncertainty from the interpretation of the dynamics of the studied CME, CME-driven shock, and background solar wind interaction responsible for the features observed in 1D and 2D.
    
    The cone CME model is being employed in scientific studies and also in operational forecasting for decades, owing to its simplicity and the ease with which its geometry can be constrained from coronagraph white-light observations \citep[e.g., ][]{2013Millward,2017Harrison,2023Verbeke}. Despite its simplicity and the absence of an internal magnetic field, \citet{2024Rodriguez} found it to be presently the model that provides the most accurate predictions for the CME arrival time at Earth. In this study, we also employed the cone CME model in order to put our work in the same context as numerous studies performed during the last few decades. 
    
\begin{figure*}[h!]
                \centering
                \includegraphics[width=0.75\textwidth]{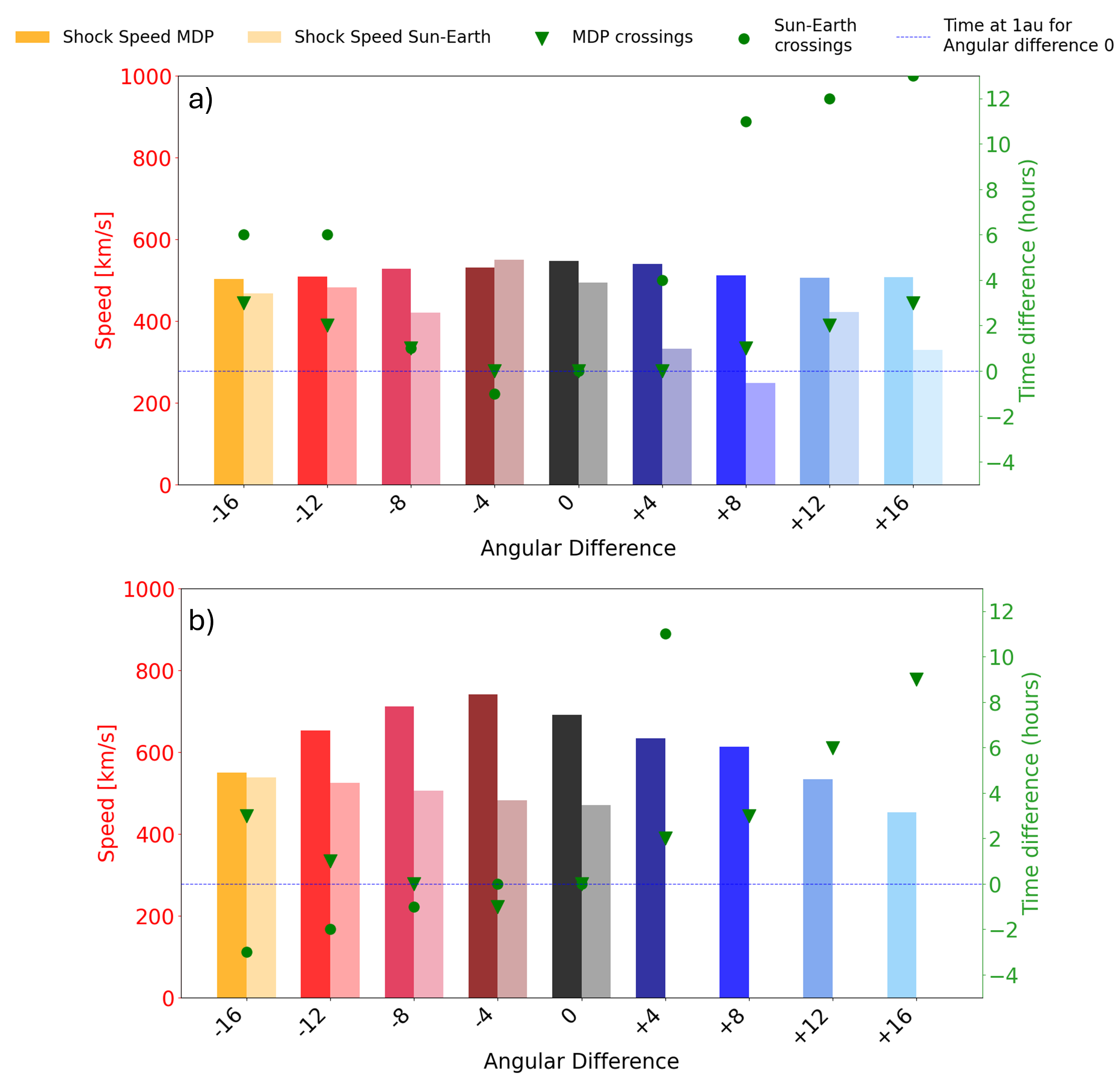}
                \caption{Time differences in CME-driven shock arrival relative to the main direction of propagation (MDP) and the Sun–Earth line. Panel~a corresponds to Event~1 and panel~b to Event~2. Dark (light) red bars indicate the shock speed at 1~au for different angular differences from the SE (MDP) line. Blue crosses (triangles) indicate the corresponding arrival time differences relative to the Sun-Earth (MDP) line.}
                \label{Fig:Time_differentce}
            \end{figure*}

    In order to trace the CME-driven shock structure and the deformations it undergoes (Figs.~\ref{Fig:Dec20_Delta_distance} and \ref{Fig:Oct21_Delta_distance}) we employed the Rankine–Hugoniot conditions (Sect.~\ref{sect:Shock_tracing_method}). Both studied events were observed as flank encounters in the in situ measurements owing to their MDP not being in the ecliptic plane \citep{2024Valentino}. In Fig.~\ref{Fig:Time_differentce} we tried to quantify the variations of the CME-driven shock arrival time at Earth, depending on the spectrum of different angles relative to both the MDP and the Sun–Earth line. 
    For Event~1, along the Sun-Earth line, the arrival at 1~au is the earliest for -4$^{\circ}$, with a time difference of -1.4~hours with respect to the Sun-Earth line. This also coincides with the fastest CME-shock speed at this position angle. Close to this location, the CME-driven shock presents a prominent knob, with the furthest front shifted towards the MDP, as can be seen in panel~c of Fig.~\ref{Fig:Dec20_Delta_distance}. All other positions show a positive time difference, with the lines furthest from the Sun-Earth line being the ones with the most difference. In particular, +16$^{\circ}$ shows a large difference of +13~hours with respect to the Sun-Earth line. On the other hand, close to the MDP, there is not so much difference in the arrival time, and the differences are relatively symmetrical with respect to the MDP, i.e., the MDP shows the fastest shock speed with speed decreasing as the angular difference increases. This reflects in the shock wave arrival time, with the maximum time difference of $\sim$3~hours seen at $\pm$16$^{\circ}$. This behaviour of the shock front close to the MDP at 1~au can be associated with the relative symmetry of the CME-driven shock, as seen in panel~c of Fig.~\ref{Fig:Dec20_Delta_distance}.
    
    In panel~b of Fig.~\ref{Fig:Time_differentce}, the latitudes North from the Sun-Earth line (positive angular differences, in the shades of light blue) are not present as at 1~au the shock structure does not arrive at Earth (this can be seen also in panel~c of Fig.~\ref{Fig:Oct21_Delta_distance}). In the case of latitudes south from the Sun-Earth line (negative angular differences, in the shades of light red), the shock arrives earlier and demonstrates the speeds larger than the one along the Sun-Earth line, with -16$^{\circ}$ showing the biggest time difference ($\sim$3~hours) and speed ($\sim$570~km~s$^{-1}$). For Event~2, the behaviour of the shock is strongly different from that of Event~1. The absolute minimum of the time difference with respect to the MDP is at -4$^{\circ}$, and it increases relatively symmetrically as the absolute angular difference increases. The maximum time difference of $\sim$9.5~hours, is reached at +16$^{\circ}$, where the shock also has the lowest speed $\sim$420~km~s$^{-1}$. This difference can be attributed to the interaction with the solar wind stream present in the simulation pointed by the pink arrow in Fig.~\ref{Fig:2021Event_with_arrows}. 
    We list below the most important points in the study of CME-driven shocks:
    
    \begin{enumerate}
        \item[a)] the CME-driven shocks, for both events, have structured fronts, with rather sharp variations in the gas compression $r_g$ value along the deformed shock surface,
        \item[b)] the shock surface deformations are, as well as the similar CME-front deformations, induced by the complex background solar wind characteristics. The western flanks of both CMEs are relatively smoother than the eastern ones, which could be due to the Parker spiral shape of the solar wind.
        \item[c)] the arrival of the CME-driven shock wave at Earth, which is very important in the operational space weather, is strongly influenced by the shock deformation. Resulting in the difference of arrival time of up to about +16h. 
        \end{enumerate}

    The results of our study provide a new perspective on the interaction of the solar wind with the CME  and CME-driven shock wave. We show what effect the structured background solar wind can have on the evolution of the CME and its arrival time at Earth, and in particular, we demonstrate the necessity of the 3D presentation for understanding this evolution. Since presently the most accurate way of testing the modelling accuracy is comparison with the in situ time series, we propose the combined approach of employing the 3D, 2D, and 1D information. We also detailed the importance and contribution of each of them. 
    
    We have shown that in the two studied events, the interaction between the ambient solar wind and the CMEs leads to different parts of the CME-driven shock developing different characteristics, which also resulted in significantly different arrival times at Earth. This result is very important in space weather predictions, in particular for CMEs with the main propagation direction being not in the ecliptic plane and resulting in a flank encounter with Earth. Our simulation results show that along the main direction of propagation, CMEs tend to propagate less perturbed with the ambient solar wind than in the case of the flank regions. In the context of space weather predictions, flank regions show the largest variation in the time of arrival of the CME to 1~au (in the order of tens of hours). It needs to be noted that the arrival time at 1~au of the CME-driven shock at the flank region can be very similar to the arrival time of the shock along the MDP if that part of the CME and CME-driven shock is interacting with the fast solar wind flow. Therefore, the accurate solar wind representation is of great importance also for the accurate forecasting of the CME and CME-driven shock wave arrival time at Earth on any other place in the heliosphere. It should also be noted that the shape of the HCS, which reflects the complexity of the global magnetic field of the Sun, is also very important for the accurate modelling of the CME and CME-driven shock arrival time at 1~au. As a further step towards better understanding of the propagation of the CMEs and CME-driven shock waves, we aim to employ magnetised CME models, like the FRi3D \citep{2016Isavnin,2022Maharana,2025Flossie} or Horseshoe \citep{2024Linan,2024Maharana}. We expect such, even more advanced modelling setup will, show how the interaction of the background solar wind and CMEs magnetic field influence the evolution of the CME and CME-driven shock and their arrival time at any place in space and time in the inner heliosphere, and make meaningful strides towards understanding the shape of these structures \citep[see e.g.][]{2025Al-Haddad}

\begin{acknowledgements}
Computational resources and services used in this work were provided by the VSC (Flemish Supercomputer Centre), funded by the FWO and the Flemish Government-Department EWI.
J. M. acknowledges the financial support by the BRAIN.be project SWIM, University start-up grant 3E220031 and the FEDtWIN project PERIHELION. J.M. and A.V. acknowledge the financial support of the University C1-project C16/24/010, UnderRadioSun.
\end{acknowledgements}

\bibliography{Bibliography}{}
\bibliographystyle{aa}

\end{document}